\documentclass[floatfix,aps,prd,abstract,twocolumn,10pt,nofootinbib,preprintnumbers]{revtex4}

\usepackage{amsgen,amsmath,amstext,amsbsy,amsopn,amssymb}
\usepackage{graphicx}
\usepackage[usenames]{color}
\usepackage{epsfig}
\usepackage{multirow}
\usepackage{longtable}

\newcommand{\mb}{\mathbf}

\newcommand{\Ddel}{\delta_{\rm D}   }

\newcommand{\MpcOh}{ \,  h^{-1}\mathrm{Mpc}}

\newcommand{\nn}{ \nonumber }
\newcommand{\Msun}{ \,   {h}^{-1}M_{\odot} }

\newcommand{\beq}{\begin{equation}}
\newcommand{\eeq}{\end{equation}}
\newcommand{\beqa}{\begin{eqnarray}}
\newcommand{\eeqa}{\end{eqnarray}}

\newcommand{\comment}[1]{}

\begin{document}

%\preprint{Version 0.99}

% unknown references     Despalietal2015, DespaliSheth2015

\title{ Effective Window Function for Lagrangian Halos}

\author{Kwan Chuen Chan$^{(1,2,3)}$} \email{chan@ice.cat}  
\author{Ravi K. Sheth$^{(4)}$} 
\author{Rom\'an Scoccimarro$^{ (5) }$} 

\affiliation{$^{1}$ D\'epartement de Physique Th\'eorique and Center for Astroparticle Physics,
Universit\'e de Gen\`eve, 24 quai Ernest Ansermet, CH--1211 Gen\`eve 4,
Switzerland}

\affiliation{$^{2}$  Institute of Space Sciences, IEEC-CSIC, Campus UAB, Carrer de Can Magrans, s/n,  08193 Bellaterra, Barcelona, Spain}

\affiliation{$^{3}$School of Physics and Astronomy, Sun Yat-Sen University, Guangzhou 510275, China }

\affiliation{$^{4}$Center for Particle Cosmology, University of Pennsylvania, 209 S. 33rd St.,  Philadelphia 19104, Pennsylvania,  USA}

\affiliation{$^{5}$Center for Cosmology and Particle Physics, Department of Physics, \\
 New York University, New York 10003, New York, USA}

\date{\today}

\begin{abstract}
  The window function for protohalos in Lagrangian space is often assumed to be a tophat in real space. We measure this profile directly and find that it is more extended than a tophat but less extended than a Gaussian; its shape is well-described by rounding the edges of the tophat by convolution with a Gaussian that has a scale length about 5 times smaller.  This effective window $W_{\rm eff}$ is particularly simple in Fourier space, and has an analytic form in real space.  Together with the excursion set bias parameters, $W_{\rm eff}$ describes the scale-dependence of the Lagrangian halo-matter cross correlation up to $kR_{\rm Lag} \sim 10 $, where $R_{\rm Lag}$ is the Lagrangian size of the protohalo.  Moreover, with this $W_{\rm eff}$, all the spectral moments of the power spectrum are finite, allowing a straightforward estimate of the excursion set peak mass function.  This estimate requires a prescription of the critical overdensity enclosed within a protohalo if it is to collapse, which we calibrate from simulations.  We find that the resulting estimate of halo abundances is only accurate to about 20\%, and we discuss why:  A tophat in `infall time' towards the protohalo center need  not correspond to a tophat in the initial spatial distribution, so models in which infall rather than smoothed overdensity is the relevant variable may be more accurate. 
\end{abstract}

\maketitle

%% Intro:
%%  Window is everywhere in biased tracer models
%%  Window is neither TH nor G.

%% Section 2:
%%  nh(<r)/ntot(<r) as real-space smoothing window applied to deltaLag
%%   -Measurements showing window is neither TH nor G, but Weff = Wth*Wg is OK
%%    -Analytic expression in real-space
%%   -Similar story from Pcross(k), xicross(r)
%%    -Dependence on concentration
%%     -Relevance for halo model decompositions?

%% Section 3:
%%  nESP with Weff
%%  biasESP with Weff in terms of b100 b010 b001 but emphasis on Weff

%% Appendix:
%%   -Dependence on linking length

\section{Introduction}
The abundance and clustering of virialized dark matter halos can be used to constrain cosmological parameters \cite{PressSchechter1974,ShethTormen1999}. The most widely studied models for halos in the late time Eulerian field are said to be Lagrangian:  one postulates that the physics of Eulerian halo formation may be understood by studying the Lagrangian protohalo patches from which they formed.  The reason for doing this is that statistics in the Lagrangian space are simpler than in Eulerian space:  this is especially true if the initial field is Gaussian.  However, defining the proto-halos theoretically becomes a non-trivial task. One has to determine what conditions must be satisfied for a Lagrangian patch to form a halo. 

In the excursion set approach \cite{BCEK1991,MussoSheth2012} the Lagrangian protohalo patches from which halos later form are modeled by requiring that the smoothed dark matter density field satisfy certain constraints.  This is also true for more elaborate models based on peaks \cite{BBKS86,BondMyers1996,Desjacques2008,DesjacquesSheth2010,DesjacquesCrocceetal2010,Desjacques2012}, and excursion set peaks \cite{AppelJones1990,ParanjapeSheth2012,ParanjapeShethDesjacques2013,BiagettiChanetal_2014}.  The smoothing window is often assumed to be a tophat in real space, based on physical grounds.  E.g., the spherical collapse model is explicitly about tophat spherical shells: the evolution of a shell is governed by the mean tophat overdensity within it \cite{GunnGott1972,Peebles1980,Padmanabhan1993}.  The tophat assumption has the added simplicity that the tophat profile is maintained during collapse, until the final violent relaxation phase \cite{LyndenBell1967}.  
%Thus tophat windows are often invoked in theoretical modeling of the Lagrangian halos due to its simplicity. 

However, there are scattered hints from previous numerical studies that the Lagrangian shape is more extended than a tophat \cite{PorcianiDekelHoffman2002,DalalWhite_etal2008,EliaLudlowPorciani2012, DespaliTormenSheth2013,Baldauf_RSDworkshop,BaldaufDesjacquesSeljak2014,Chan2015}.  Moreover, the peaks-based predictions for halo abundance and formation require integrals (over the initial power spectrum), some of which are ill-defined for a tophat window.  To alleviate this problem, it is common to use a Gaussian window instead \cite{BBKS86}, or to use a tophat when it leads to convergent results and a Gaussian otherwise \citep[e.g.][]{ParanjapeShethDesjacques2013, ParanjapeSefusattietal_2013, BiagettiChanetal_2014}.  Finding a better motivated, less ad hoc treatment is desirable.  

In these approaches, the window function affects both the predicted abundance of halos, as well as their clustering.  On large scales, the clustering of halos is linearly biased with respect to the matter.  As measurements and theoretical predictions become more precise, it has become necessary to account for the fact that this bias may be scale dependent \cite[see][for a recent review]{Desjacques:2016bnm}.  Correctly modeling this scale dependence requires a good understanding of the window function \cite{BBKS86,MussoParanjapeSheth2012,LagBiasPaper}.  Thus, both when predicting halo abundances and when modeling halo bias, the shape of the window function plays a crucial role.  The main goal of the present study is to use numerical simulations to determine this shape.  

%In these models the spectral moments of the power spectrum are essential, however, high order spectral moments are ill-defined for tophat window due to its slow decay in Fourier space.  To alleviate this problem, one usually takes a hybrid approach by computing the zeroth moment with a tophat, while higher order ones with a Gaussian window, e.g.~\cite{ParanjapeShethDesjacques2013, ParanjapeSefusattietal_2013, BiagettiChanetal_2014}.  It is more satisfactory if one can find a unified treatment for all the moments.  Another important reason for modeling the window function well is that these models predict the bias parameters to be scale-dependent. While they predict the scale-dependent bias parameters in precise form, the window function is left unconstrained.  As both the window function and the bias parameters give rise to scale-dependence, one must model both of them well in order to disentangle their scale-dependent contributions. 

%This paper is organized as follows.
In Sec.~\ref{sec:Lagr_profile_realspace}, we describe real-space estimates of the effective window $W_{\rm eff}$ from simulations, and propose a simple analytic parametrization of it. Fourier space estimates are the subject of Sec.~\ref{sec:Lag_cross_bias}. In Sec.~\ref{sec:nesp}, we combine $W_{\rm eff}$ with the collapse threshold measured directly in simulations to see if the excursion set peak approach correctly predicts the halo mass function. We conclude in Sec.~\ref{sec:conclusion}.  Several details are provided in Appendices.  Appendix~\ref{sec:sys} describes a number of systematics: discreteness effects, estimating $W_{\rm eff}$ from a clustered field rather than the initial Lagrangian grid, and the effect of fixing some parameters when fitting the protohalo-matter cross bias parameter.  Dependence on how the halos were defined in the first place is studied in Appendix~\ref{sec:hf}.  Appendix~\ref{sec:sc} discusses how $W_{\rm eff}$ can be understood in the context of the spherical collapse model.

\section{ Lagrangian Window Function }
\label{sec:Lagr_profile_realspace}

Suppose that one has identified halos in the late time Eulerian field.  Then, using the initial positions of the particles, one defines the Lagrangian protohalo patch as the region from which each Eulerian halo forms.  In Lagrangian models for halos, one postulates that the physics of Eulerian halo formation may be understood by studying the Lagrangian protohalos.  The reason for doing this is that statistics in the Lagrangian space are simpler than that in Eulerian space:  this is especially true if the initial field is Gaussian.  The cost -- there is always a price to pay -- is the complexity required to define a protohalo in Lagrangian space (the definition is usually relatively simple in Eulerian space).  E.g., are protohalos peaks in density? If so on what scale? Is density the only variable which matters?  Etc.

In this work, we use two sets of simulations from the LasDamas project: Oriana and Carmen. Both simulation sets assume the same flat $\Lambda$CDM model with cosmological parameters $\Omega_{\rm m} = 0.25$, $\Omega_{\Lambda} = 0.75$ and $\sigma_8 =0.8$.  The transfer function is taken from CMBFAST \cite{CMBFAST}. The initial conditions are Gaussian with spectral index $n_s=1$.   The initial particle displacement fields are set using 2LPT \cite{CroccePeublasetal2006} at $z_*=49$, after which the particles are evolved using the public code Gadget2 \cite{Gadget2}.  In the Oriana simulations, there are $1280^3$ particles in a cubic box of size 2400 $ \MpcOh $, and $1120^3 $ particles in a box of size 1000 $ \MpcOh $ in the Carmen simulations.   Thus, in the Oriana and Carmen simulations, each particle carries a mass of $4.57  \times 10^{11} $ and  $4.94  \times 10^{10} h^{-1}M_{\odot}$ respectively.  Our results are averaged over 5 realizations for Oriana, and 7 for Carmen.  In each simulation, the Eulerian halos are identified at $z=0.97$ and $z=0$ using the Friends-Of-Friends algorithm \cite[][hereafter FOF]{Davisetal1985} with linking length $\ell=0.156$ times the interparticle separation. We show results for $\ell=0.2$ in Appendix~\ref{sec:ll_dependence}.  To resolve the halo profiles well, we only consider halos with at least 60 particles; we discuss discreteness effects in Appendix~\ref{sec:discrete}.  We bin the halos into narrow mass bins of width $\Delta \ln M = 0.157$.  

%To study the protohalos in Lagrangian space numerically, we proceed as follows.
For each Eulerian halo identified at redshift $z$ (typically $z=0$ and 0.97 in this paper) we trace back its constituent particles to the initial redshift $z_*$. The center of mass of the corresponding Lagrangian protohalo is estimated using the center of mass of the constituent particles in the initial Lagrangian space.  If the protohalo patches were spherical, then the range of masses in a bin corresponds to a range of $\sim 5\%$ in Lagrangian radii.  This is sufficiently small that we do not expect the results which follow to be affected by the binning.

%%%%%%%%%%%%%%%%%%%%%%%%%%%%%%%%%%%%%%%%%%%%%%%%%%%%%%%%%%%%%%%%%%%%%%%%%%%
\subsection{The Lagrangian window in real space}
\label{sec:LagWin_real}
We expect the window function $\mathcal{W}$ to have a number of properties. First, to define a halo in Lagrangian space, we expect it to be compact and reasonably well localized in real space.  We use $R$ to denote the `scale' of the filter, where 
%also require $ W(\mb{x}) \geq 0 $ and 
\beq
 \label{eq:W_integralconstraint}
 \int d\mb{x}\, \mathcal{W}_{\rm R}(\mb{x}) \equiv V_{\rm R} \propto R^3. 
 \eeq
 (i.e.~$\mathcal{W}_{\rm R} $ is normalized such that $ \mathcal{W}_{\rm R}(0)=1 $).  Although we know that halos, and the protohalo patches from which they formed, are not really spherical \citep[e.g.][]{DespaliTormenSheth2013,Despalietal2017}, we will assume, for simplicity, that $\mathcal{W}$ is spherically symmetric.  With these minimal conditions, we now describe how we reconstruct the effective window function from the Lagrangian halo profile.

Consider a protohalo centered at the origin.  The total number of particles in this halo, $N$, is given by 
\beq
\label{eq:N_pnm}
N  =  \int d\mb{x}  \,  n_{\rm h} ( \mb{x} )
%   =  \int d\mb{x}  \,  p_{\rm h} ( \mb{x} ) \, n_{\rm m} ( \mb{x} ), 
   =  \int dr\,4\pi r^2 \, n_{\rm m}(r)\,  p_{\rm h}(r), 
\eeq
where $ n_{\rm m } $ is the number density of dark matter particles, and $p_{\rm h}$ is the probability that a dark matter particle at distance $r$ from the protohalo center is part of the protohalo.  This shows that we can estimate $p_{\rm h}(r)$ for each protohalo from the ratio  $ n_{\rm h}(r) /  n_{\rm m}(r) $, where $ n_{\rm h}(r) $ is the number density of particles belonging to the protohalo that are at distance $r$ from the protohalo center.  When the dark matter particle distribution is uniform, as it is in the initial conditions, and discreteness effects are not a concern, then $n_{\rm m}(r)=\bar n_{\rm m}$ and so $N=\bar n_{\rm m}\,V_{\rm R}$.

\begin{figure*}[!htb]
\centering
\includegraphics[width=\linewidth]{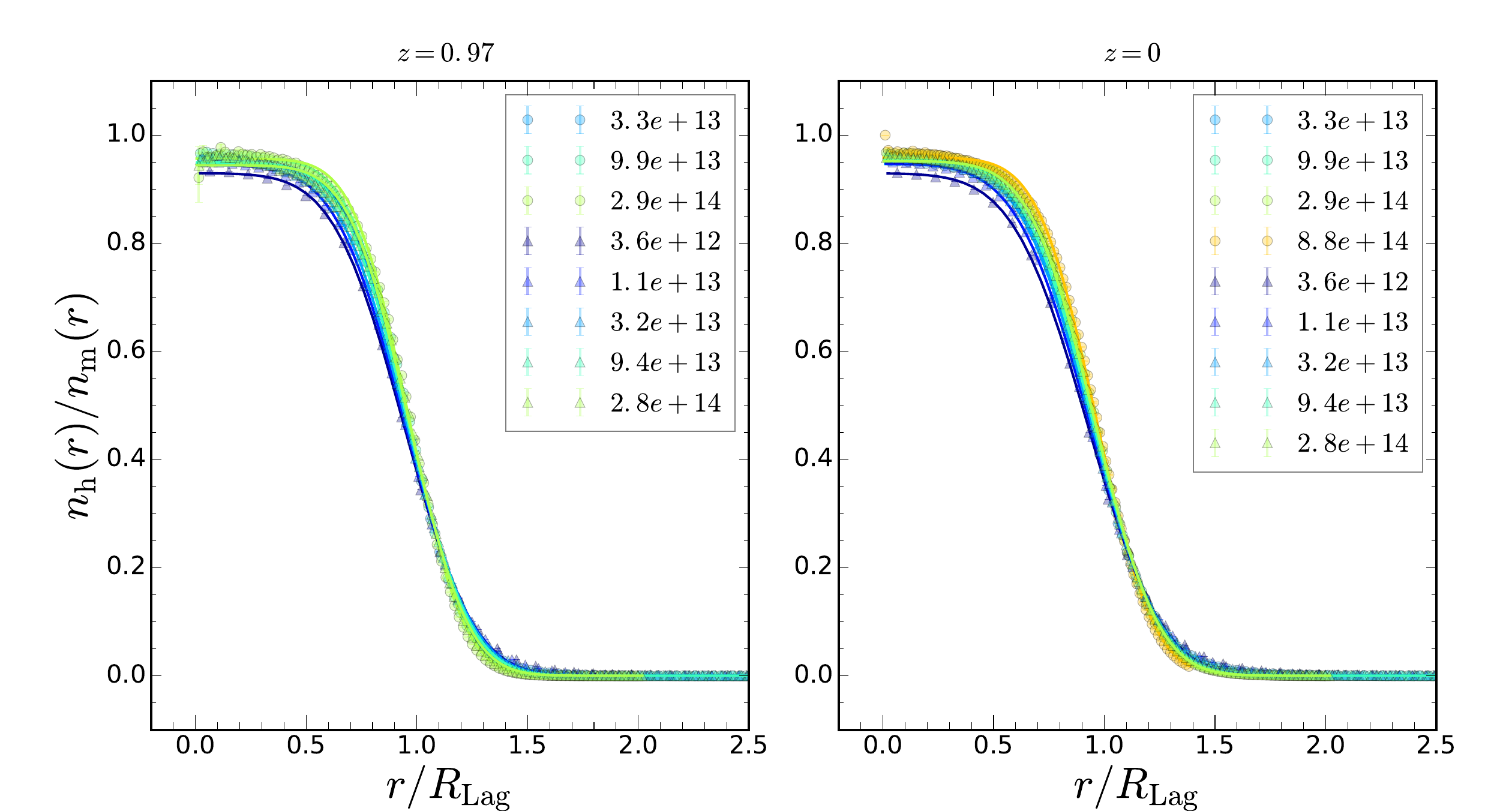}
\caption{ The estimator $n_{\rm h} / n_{\rm m} $ for  $p_{\rm h}$ in Oriana (circles) and Carmen (triangles) for the range of halo masses shown in the legend (in units of $ \Msun$). Solid lines, same color as the symbols, show the result of fitting Eq.~\eqref{eq:WG*THrealspace} to the measurements.}
\label{fig:nh_ntot_ratio}
\end{figure*}

%$n_{\rm h}$ is the number density of dark matter particles which belong to it. Let the probability that a dark matter particle at a distance $r$ is incorporated into the Lagrangian halo be  $p_{\rm h}$, then we can write $N$ as 
%\beq
%\label{eq:N_pnm}
%N  =  \int d r\, 4 \pi r^2\, p_{\rm h}(r)\,  n_{\rm m } ( r ),   
%\eeq  
%where $ n_{\rm m } $ is the number density of the underlying dark matter density, which is different from the mean dark matter number density $ \bar{ n}_{ \rm m} $ if the clustering is not negligible. To construct the Lagrangian halo profile, we will remove the clustering effect either by tracing the particles back to their initial grid points or by modeling the clustering and then dividing it out.  We will show results from both approaches. As we assume that the clustering effects will be removed by either means, we can safely take  $n_{\rm m } $ to be  $ \bar{ n}_{ \rm m} $ here. 

Even when the Lagrangian halo profile is not tophat, it is convenient to define the Lagrangian radius $R_{\rm Lag}$ from 
\beq
 N \equiv \frac{  4 \pi }{ 3 } \bar{n}_{ \rm m} R_{ \rm Lag}^3 ,
\eeq
where $\bar{n}_{ \rm m}$ is the mean density of dark matter particles.  Then Eq.~\eqref{eq:W_integralconstraint} indicates that we should define 
\beq
 %W (r) \equiv \frac{  3\, p_{\rm h}(r)\, n_{ \rm m}(r)/\bar{n}_{ \rm m}}{  4 \pi  R_{\rm Lag}^3 }.
 \frac{ \mathcal{W} (r)}{V_{\rm R}} \equiv \frac{  3\, p_{\rm h}(r) }{  4 \pi  R_{\rm Lag}^3 } \frac{ n_{ \rm m}(r)} { \bar{n}_{ \rm m } }.
\eeq
For the rest of the paper, we follow the convention of \cite{BBKS86} and use  $W \equiv   \mathcal{W} / V_{\rm R}$, which is of dimension 1/Volume in real space and dimensionless in Fourier space.  In the initial conditions, $ W(r) = p_{\rm h}(r)/(4 \pi  R_{\rm Lag}^3/3)$.  This shows that $p(r)$ is proportional to the shape of the window.

%%  Because of the presence of the halo, the dark matter clustering around its center is enhanced by a factor of $1 + \xi_{\rm c} $, where $ \xi_{\rm c} $ is the cross correlation function between halo and matter. In terms of the mean dark matter number density $\bar{n}_{\rm m} $,  $N$ can be expressed as 
%% \beq
%% N  =  \int d x' 4 \pi x'^2  W( \mb{x}' ) \bar{n}_{\rm m} [ 1 + \xi_{\rm c} ( \mb{x}' ) ].  
%% \eeq
%% Therefore we can estimate the window $W$ as
%% \beqa
%% \label{eq:W_estimator1}
%% W(r) &=& \frac{ n_{\rm h}(r) }{ n_{\rm tot}(r) }  \\
%% \label{eq:W_estimator2}
%%  W(r) &=&  \frac{ n_{\rm h}(r) }{ \bar{n}_{\rm m}[1 + \xi_{\rm c}( r) ] } ,
%% \eeqa
%% where $ n_{\rm h }(r) $ is the number density of the particles belonging to the Lagragnian halos and residing at a distance $r$ from the center and  $ n_{\rm tot} $ is the number density of all the dark matter particles at a distance $r$ from the center.  Eq.~\eqref{eq:W_estimator1} and \ref{eq:W_estimator2} represent two different ways to estimate  $W$, and we will discuss both of them. 

We estimate $p_{\rm h}(r)$ for each protohalo from the ratio  $ n_{\rm h}(r) /  n_{\rm m}(r) $, where $ n_{\rm h}(r) $ is the number density of particles belonging to the Lagrangian halo.  The results are shown in Fig.~\ref{fig:nh_ntot_ratio}. To estimate $ n_{\rm h}(r)$, for each Lagrangian halo, we bin all the protohalo particles belonging to the Lagrangian halo into spherical shells about the center of mass of the halo, and we then compute the number density in each shell.  The spherically averaged number density is further averaged over all the halos in the same mass bin.  We estimate $n_{\rm m}(r)$ by carrying out the same procedure for all the particles (protohalo or not).  In both cases, we use the particles traced back to the initial grid points, and not the 2LPT displaced positions.  See Appendix~\ref{sec:displaced} for results based on the 2LPT positions.

Fig.~\ref{fig:nh_ntot_ratio} shows the measurements from both Carmen and Oriana for the Lagrangian protohalos of Eulerian halos identified at $z=0.97$ and 0 respectively. The results from the two redshifts are very similar.  Normalizing distances by the mean $R_{\rm Lag}$ in the mass bin (rather than doing this halo by halo) removes most of the dependence on mass.  Indeed, in this normalized variable, protohalos spanning three orders of magnitude in mass follow very similar profiles.  However, there is  a small residual mass dependence:  the less massive protohalos are slightly more diffuse than the more massive ones.  In general, $p_{\rm h}\sim 1$ on small scales, but when $r/R_{\rm Lag}\ge 1$, then the profile drops sharply: it is smoother than a tophat and more localized than a Gaussian, in qualitative agreement with Fig.~3 of \cite{DalalWhite_etal2008}.

The careful reader will have noted that $p_{\rm h}$ differs from unity at small $r$.  Appendix~\ref{sec:discrete} discusses this limit and shows that some of the difference from unity is a consequence of discreteness effects, especially for the least massive halos.  Moreover, if the protohalo consists of more than one lump, and we use only the main lump in the Lagrangian protohalo as in \cite{DespaliTormenSheth2013}, the resultant $ p_{\rm h} $ near the center is closer to unity.  Thus, while it is possible that a small fraction of particles from the center of the protohalo escape, this fraction is smaller than the Figure indicates.  

\subsection{Fitting function for the Lagrangian window}
We noted above that $p_{\rm h}(r)$ is proportional to a smoothing window.  For a generic smoothing window, the value of the smoothed overdensity fluctuation field at position $\mb{x}$ is 
\begin{align}
  \delta_{\rm W}(\mb{x} ) &= \int d \mb{x}'\, W( \mb{x} - \mb{x}' )\, \delta_{\rm m}(\mb{x}')        \nn \\
%    &= \frac{1}{ \bar{\rho}_{\rm m} }  \int d \mb{x}' W( \mb{x} - \mb{x}' ) \rho_{\rm m}( \mb{x}' )   - 1 \nn \\
   &= \frac{1}{ \bar{n}_{\rm m} }  \int d \mb{x}'\, W( \mb{x} - \mb{x}')\, n_{\rm m}( \mb{x}') - 1 .
\end{align}
This shows that, when the position $\mb{x}$ corresponds to a protohalo center, then $\delta_{\rm W}(\mb{x})$ is the matter overdensity smoothed with $W\propto p_{\rm h}(r)$.  Fig.~\ref{fig:nh_ntot_ratio} shows that $W$ is rather different from (it has rounder edges than) a tophat.  

Consider a tophat of radius $R_{\rm Lag}$ centered on a protohalo patch.  There are at least three reasons why we expect $W$ to differ from this tophat.  First, the protohalos are not spherical, so assuming spherical symmetry can make $p\ne 1$.  \cite{Despalietal2017} shows that this matters, but is a small effect.  Second, particles within $R_{\rm Lag}$ having velocities larger than the escape speed will escape (i.e. move beyond $R_{\rm Lag}$); others which were initially beyond $R_{\rm Lag}$ may have had velocities which brought them closer.  Appendix~\ref{sec:fuzzy} discusses why both types of particles will tend to smear out the sharp edge.  And third, the center of mass of the protohalo patch may be a good but not perfect indicator of the position around which the main collapse occurs; if so, then using it will provide a slightly smeared-out version of the `true' position of the protohalo, and hence of the true $p_{\rm h}$ shape.

The discussion above motivates us to consider a composite window consisting of a tophat smoothed by a Gaussian.  %In principle, the mean and width of this Gaussian depend on distance from the protohalo center.  In what follows, we ignore this subtlety.
I.e., we define 
\beqa
\label{eq:Weff}
 W_{\rm eff}(kR_{\rm Lag}) &\equiv& W_{\rm TH}( k R_{\rm Lag} ) \,
          W_{\rm G} \big(  \frac{k R_{\rm Lag} }{\phi} \big), 
\eeqa
where 
\beq
 W_{\rm TH} ( x) \equiv  \frac{3 }{ x^3 } ( \sin x - x \cos x ) 
 \ {\rm and}\
 W_{\rm G} (x )  \equiv   e^{ -x^2/2 } 
 \eeq
and $\phi\equiv 5/\sqrt{f}$,
%$W_{\rm TH}$ and $W_{\rm G}$ denote the tophat and Gaussian window functions we defined earlier.  
where the factor of 5 is because, as we will see shortly, the scale of the Gaussian window is about 1/5 of the size of the tophat ($R_{\rm Lag}$); in turn, this happens to be approximately the same as the Eulerian scale.  The parameter $f$ quantifies how the scale of $W_{\rm G}$ differs from $R_{\rm Lag}/5$:  in particular, $W_{\rm eff}\to W_{\rm TH}$ as $f\to 0$ whereas $W_{\rm eff}\to W_{\rm G}$ as $f\to\infty$.  I.e., larger $f$ values imply more smearing-out.

%% \begin{figure*}[!htb]
%% \centering
%% \includegraphics[width=\linewidth]{LagrHalo_OrianaCarmen_WGWTH_Rnorm_sub.pdf}
%% \caption{ The Lagrangian halo profile at $z_*$ from Carmen (circular data points) and Oriana (triangular data points) simulations, obtained from Eulerian halos at $z=0.97$ (left panel) and 0 (right panel). The mass of the halos are shown in the legend (in unit of $ \Msun $).  The best fit using  $W_{\rm  G*TH} $ window is also shown.  The model fits the numerical profile well.   }
%% \label{fig:LagrHalo_OrianaCarmen_WGWTH_Rnorm_sub}
%% \end{figure*}

One of the virtues of this functional form is that its inverse Fourier transform is analytic.  The effective window in real space is 
%% \beq
%% W_{ \rm G * TH  } ( r ) = \int \frac{d k}{ (2 \pi)^3  } 4 \pi k^2 \frac{\sin kr }{ kr } W_{\rm TH} ( kR ) W_{\rm G} ( \frac{\sqrt{f }  kR }{5} ) ,
%% \eeq
%% where we have introduced a factor of $\sqrt{f} $ in the exponential so that $R_{\rm TH}$  and $R_{\rm G} $ can vary independently. This integral can be done analytically, and we find
\beqa
\label{eq:WG*THrealspace}
W_{ \rm eff } ( r )  &=& \frac{\mathcal{W}_{ \rm eff } ( r )}{V_{\rm eff}}  =  \frac{ 3 }{ 4\pi\, R_{\rm Lag}^3 } 
    \Bigg[ \frac{e^{- x_+^2 } - e^{- x_-^2 } }{ \phi\,x \sqrt{ 2 \pi } } \nn \\
       & &   \quad  + \frac{\mathrm{erf}( x_+ )}{2} 
    -  \frac{\mathrm{erf}(| x_{-} | )}{2}  \frac{x_-}{|x_-|}  \,   \Bigg],  
\eeqa
where 
\beq
 x\equiv \frac{r}{R_{\rm Lag}},\quad
 x_\pm \equiv \frac{ \phi( x \pm 1)}{ \sqrt{ 2}}, \quad
 %x_{-} \equiv \frac{ y( x - 1)}{ \sqrt{ 2}}, \ \ 
 \phi\equiv \frac{5}{\sqrt{f}}, 
\eeq
%and erf denotes the sign and error functions respectively. 
and
\beq
 V_{\rm eff} =  V_{\rm TH}
  \left[{\rm erf}\left(\frac{\phi}{\sqrt{2}}\right) - 2\phi\, \frac{{\rm e}^{-\phi^2/2}}{\sqrt{2\pi}}\right]^{-1}  .
  \eeq
  Note that $V_{\rm eff}$ interpolates between $V_{\rm TH}\equiv 4\pi R_{\rm Lag}^3/3$ when $f\ll 1$ and grows to $V_{\rm G}\equiv (2\pi\, f)^{3/2}  \, (R_{\rm TH}/ 5 )^{3}$ when $f\gg 1$. 
 If $f< 2$  ($f<4$), then $V_{\rm eff}$ is one (ten) percent larger than $V_{\rm TH}$.  This will be important in what follows.  

%We also show the best fit using the window function Eq.~\eqref{eq:WG*TH_realspace} in  Fig.~\ref{fig:nh_ntot_ratio}.  To be general we allow additional fitting parameters to vary. More precisely, we consider the model 
%\beq
%\label{eq:profile_model}
% p_{\rm h} (r)  = \frac{4 \pi A R_{\rm Lag}^3  }{3}\, W_{\rm eff} ( r; f,  R ) , 
%\eeq
%and regard $R$, $A$ and $f$ as free parameters. We have replaced $R_{\rm Lag} $ in Eq.~\eqref{eq:WG*TH_realspace} by a free parameter $R$.   Overall, we find that this model yields a good fit to the Lagrangian profile for the numerical profile. However, we also note that the profile model tends to be slightly more concentrated than the numerical result.   

\begin{figure}[!htb]
\centering
\includegraphics[width=\linewidth]{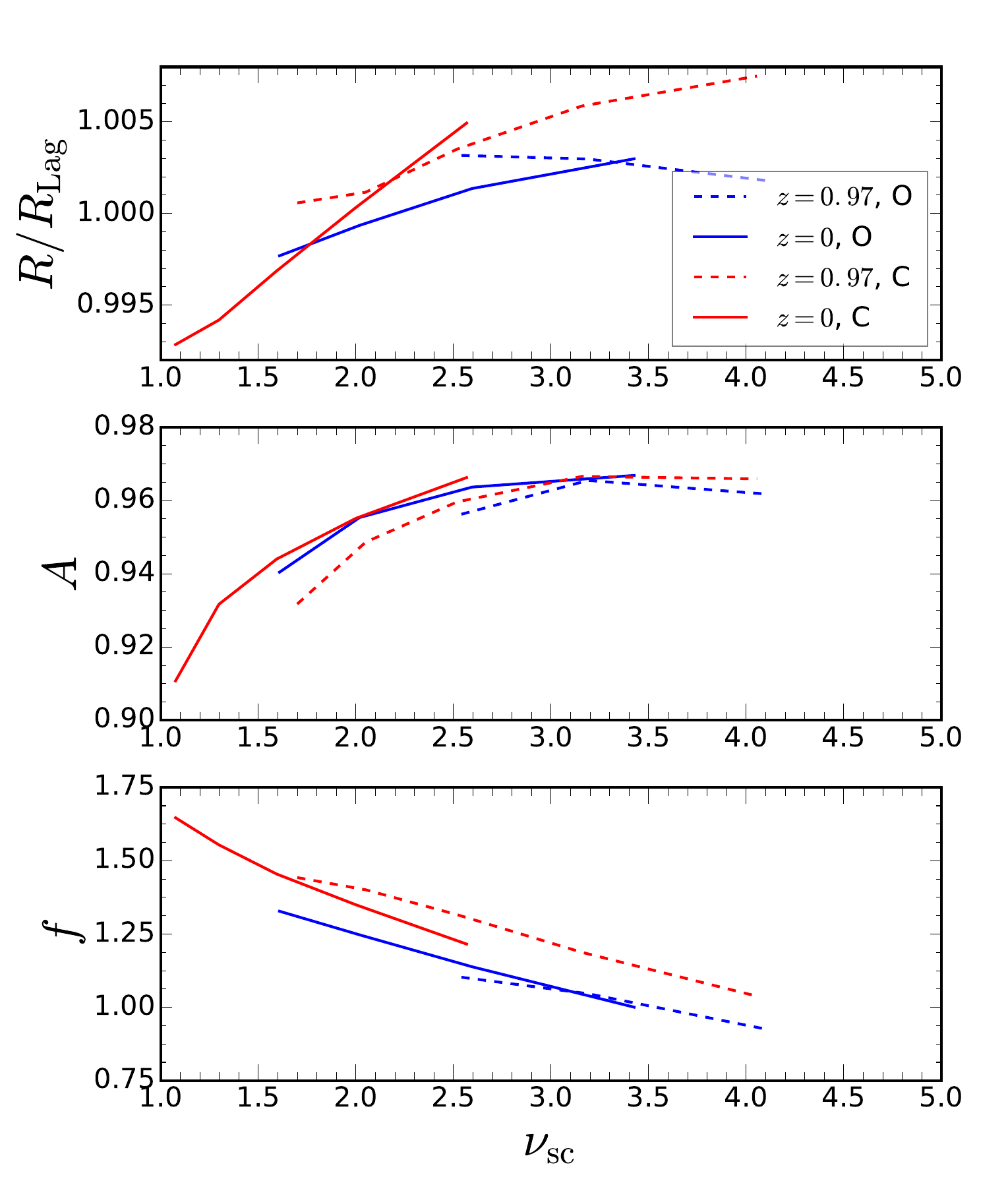}
\caption{ The best-fit parameters $\rho$, $A$ and $f$ for the Lagrangian protohalos of halos identified at $z=0.97$ (dashed) and $0$ (solid) in Oriana (blue) and Carmen (red).    }
\label{fig:LagrProfile_bestfit_WGWTH}
\end{figure}

 When fitting Eq.~\eqref{eq:WG*THrealspace} to the data in Fig.~\ref{fig:nh_ntot_ratio} we replace $R_{\rm Lag}\to\rho R_{\rm Lag}$, and allow $\rho$ and $f$ to vary.  In addition, we multiply the expression above by an overall amplitude $A$, which we also allow to vary.  Fig.~\ref{fig:LagrProfile_bestfit_WGWTH} shows the best fit parameters.  We expect $\rho$  and $A$ to be close to unity for all masses, and Fig.~\ref{fig:LagrProfile_bestfit_WGWTH} shows that this is indeed the case.  On the other hand, $f$ is a stronger function of halo mass:  it is larger than unity -- meaning that the edges of the tophat are more rounded -- at small masses, and decreases at large masses.  This mass-dependence is different for the two redshifts we have studied, but the redshift dependence is reduced significantly if we express the masses and redshifts in terms of the scaled variable
\beq
\nu_{\rm sc} \equiv  \frac{ \delta_{\rm sc} (z) }{ \sigma_0(M) },
\eeq
where $\delta_{\rm sc}(z)$ is the time-dependent spherical collapse threshold  and $\sigma_0(M)$ is defined in Eq.~\eqref{eq:sj}.   %When this is done, $f\sim 1.50$ at $\nu_{\rm sc} \sim 1.5$, but is smaller than unity at $\nu_{\rm sc} \gg 1$, passing through unity  at $\nu_{\rm sc} \sim 4 $.
When this is done, we find that when $\nu_{\rm sc} \gtrsim  3 $, the best fit $f$ is close to 1, but it increases as $\nu_{\rm sc} $ decreases.   Appendix~\ref{sec:More_on_bc_fit} shows that a Fourier space analysis returns similar results.  That $f$ depends on $\nu_{\rm sc}$ rather than mass suggests that discreteness effects (which scale with particle number) are unlikely to be the dominant cause of the smearing-out we see in Fig.~\ref{fig:nh_ntot_ratio}.

\subsection{Fourier space estimate of the window }
\label{sec:Lag_cross_bias}

\begin{figure*}
\centering
\includegraphics[width=\linewidth]{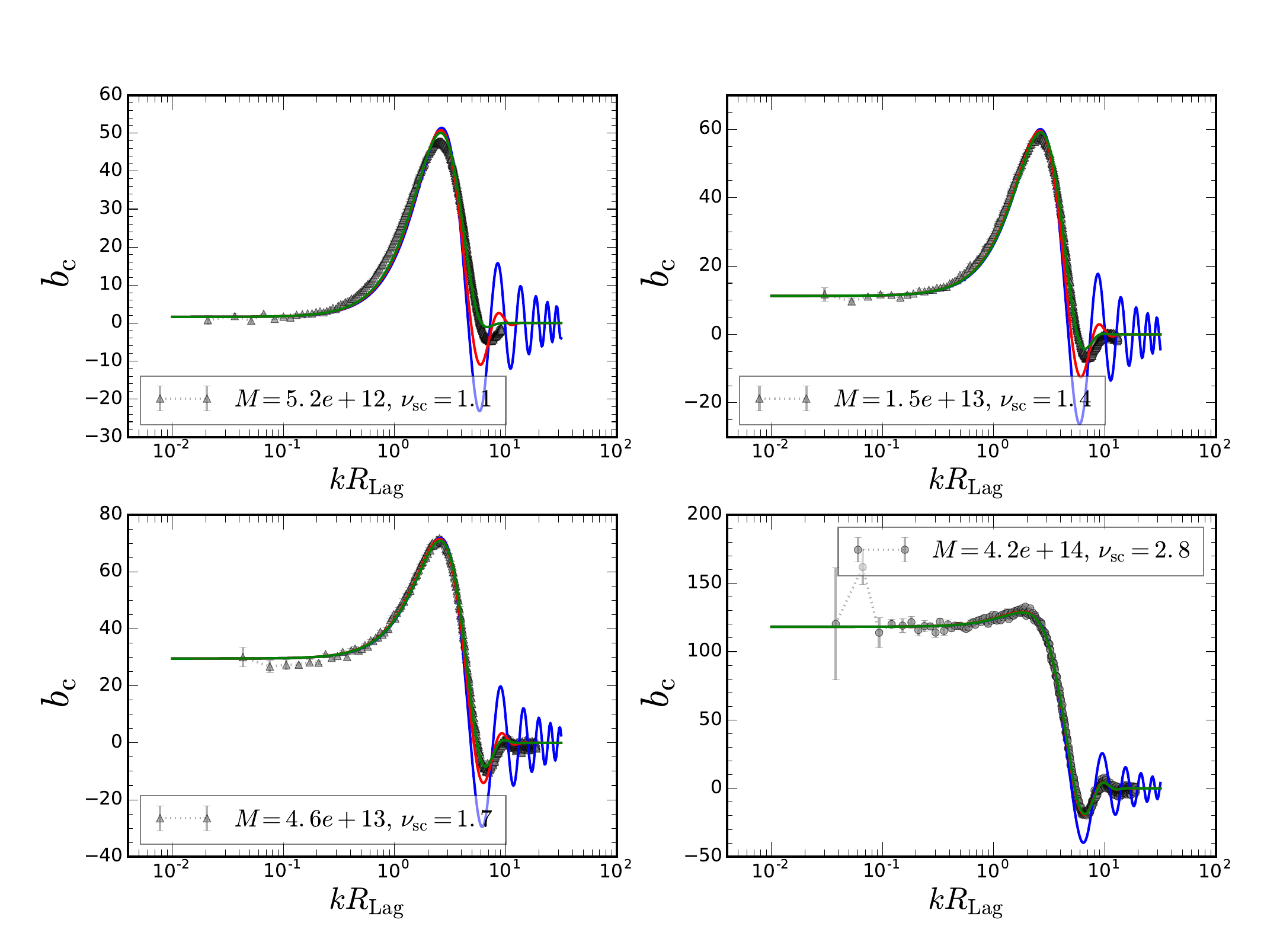}
\caption{Lagrangian cross bias parameter $b_{\rm c}$ for the Lagrangian protohalos of Eulerian halos identified at $z=0$  in the Oriana (grey circles) and Carmen (grey triangles) simulations, for a range of halo masses (legend, in units of $\Msun$).  Curves show the result of fitting Eq.~\eqref{eq:bceff} to these measurements with $W=W_{\rm TH}$ (blue) and $W_{\rm eff}$ (red), when $b_{10}$, $b_{01}$, and $R$ are varied and $f=1$.  Green shows $W_{\rm eff}$ when $f$ is also varied.  In some cases, the green and red curves overlap (e.g., when $\nu_{\rm sc} = 2.8 $). }
\label{fig:bc_TH_Eff_compare}
\end{figure*}

We now focus on the cross correlation between the protohalo centers and the dark matter; we will not consider the auto-correlation.  In Fourier space, the cross power spectrum between the Lagrangian halo density contrast, $\delta_{\rm h}$, and that of the dark matter at the initial time, $\delta_{\rm m} $, is 
\beq
 \langle \delta_{\rm h } ( \mb{k}_1 ) \delta_{\rm m}( \mb{k}_2 ) \rangle =  (2 \pi )^3  P_{\rm c } ( k_1) \Ddel( \mb{k}_{12} ) ,
\eeq
where  $\Ddel$ is the Dirac delta function.  Similarly, the Lagrangian matter power spectrum is defined as 
\beq
 \langle \delta_{\rm m}( \mb{k}_1 ) \delta_{\rm m}( \mb{k}_2 ) \rangle  =  (2 \pi )^3  P_{\rm m}( k_1) \Ddel( \mb{k}_{12} )  . 
\eeq
In what follows, $\delta_{\rm m}$ is taken from the initial Gaussian random fluctuation field. We discuss the differences when the evolved 2LPT field is used in Appendix \ref{sec:displaced}. 

The halo-matter cross-correlation function is related to the auto-correlation of the dark matter by a single multiplicative (possibly scale-dependent) bias factor \citep{FruscianteSheth2012}.  We estimate this Lagrangian cross bias parameter as 
\beq
\label{eq:bcL_k}
b_{\rm c }(k, z)  =  \frac{ D(z_{*}) }{ D(0) } \Big[ \frac{P_{\rm c}(k,z_*;z)   }{ P_{\rm m}(k,z_*)  } - 1 \Big] ,
\eeq
where we extrapolate the Lagrangian bias parameter to $z=0$ using the linear growth factor $D$.  We have included the term $-1$ because of the initial redshift of the simulation is finite (e.g.~\cite{ChanScoccimarroSheth2012}).  The argument $z$ denotes the redshift at which the Eulerian halos were identified.  

Fig.~\ref{fig:bc_TH_Eff_compare} shows our measurements of $b_{\rm c} $ (the right hand side of Eq.~\eqref{eq:bcL_k}) for the protohalo patches of halos identified at $z=0$ for a range of halo masses (as labelled in each panel).  Plotting $b_{\rm c}$ versus $kR_{\rm Lag}$ rather than $k$ alone removes some of the dependence on mass.  For $kR_{\rm Lag} \lesssim 1$, this quantity approaches a (mass-dependent) constant; it rises to a maximum at $kR_{\rm Lag}\sim 1$, after which it drops sharply and oscillates with diminishing amplitude as $kR_{\rm Lag}$ increases. We will see shortly that these oscillations have an important implication.

\begin{figure*}[!htb]
\centering
\includegraphics[width=\linewidth]{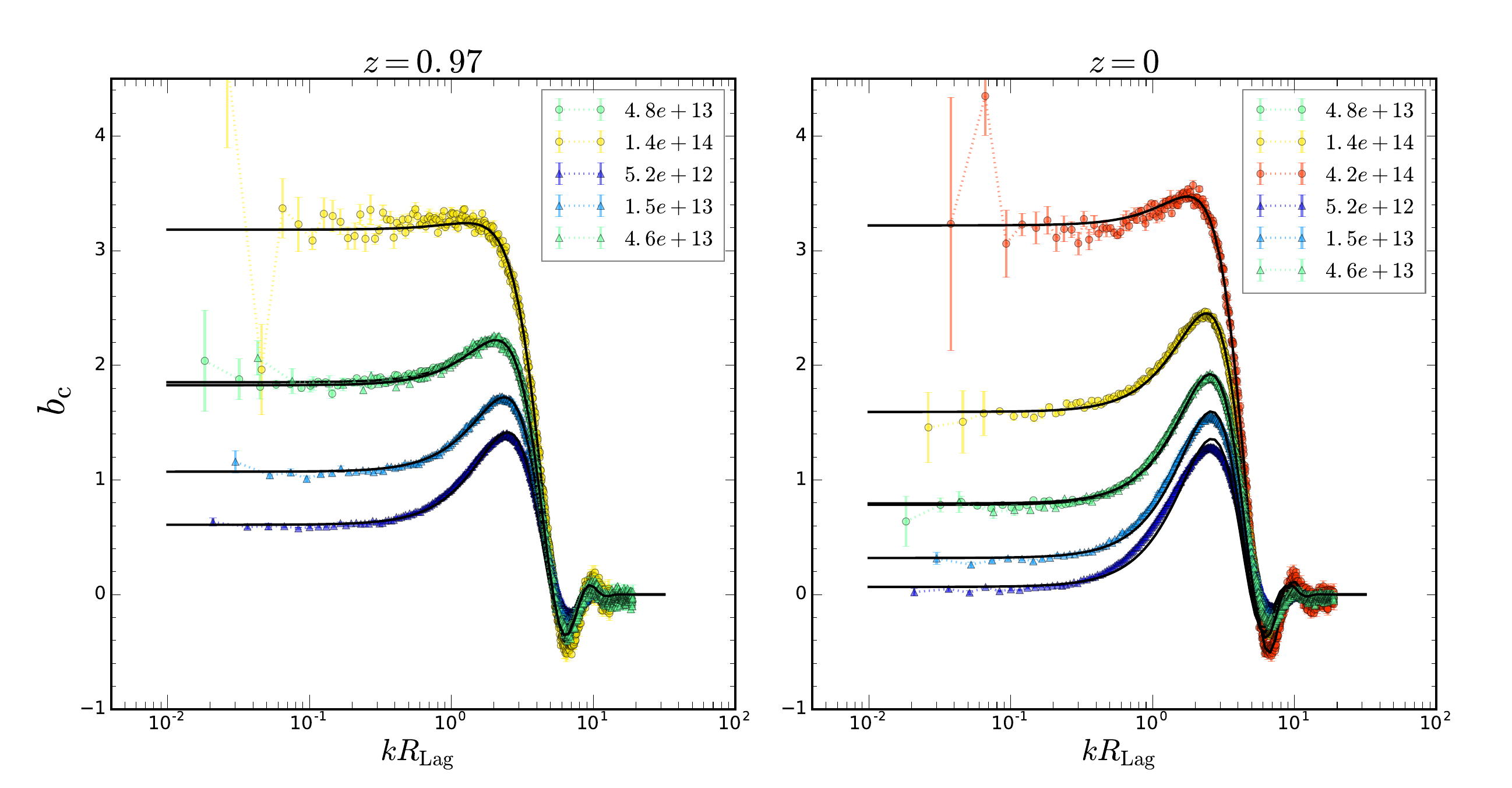}
\caption{ Lagrangian cross bias parameter $b_{\rm c}$ for the Lagrangian protohalos of Eulerian halos at $z=0.97$ (left) and $z=0$ (right) in the Oriana (circle) and Carmen (triangle) simulations, for a range of halo masses (legend, in units of $\Msun$).  Solid black curves show the result of fitting Eq.~\eqref{eq:bceff} to these measurements with $W=W_{\rm eff}$ and $f=1$ in  Eq.~\eqref{eq:Weff}.    }
\label{fig:bc_Lag_Dextrap}
\end{figure*}

In the excursion set approach, the Lagrangian protohalo-matter cross bias is \citep{MussoParanjapeSheth2012}
\beq
\label{eq:bceff} 
 b_{\rm W} (k ) = \left[b_{10}
 + 2b_{01}\,  \frac{  d \ln W(k) }{d \ln s_0 } \right]\,W(k),
\eeq
where 
\beq
\label{eq:sj}
 s_{j}  \equiv \sigma_j^2 =  \int \frac{dk}{k} \frac{ 4 \pi k^3 }{ (2 \pi)^3 } k^{2j} P_{\rm m}(k,0)\, W^2( k R ), 
\eeq
$P_{\rm m}(k,0) $ is the initial dark matter power spectrum extrapolated using linear theory to $z=0$, and $b_{10}$ and $b_{01}$ are dimensionless bias parameters whose numerical values depend on protohalo mass (i.e.~on $R_{\rm Lag}$).  Our notation highlights the fact that the smoothing window $W$ plays a crucial role.  

%These two bias parameters arise from two constraints on the excursion set trajectories. The $b_{10}$-term is due to the threshold constraint that whenever the trajectory crosses it at $s_0$, it is interpreted as the site of halo formation with halo mass associated with $s_0$.  The $b_{01}$-term comes from the first crossing constraint that the trajectory has to satisfy a further condition that it  does not cross the threshold at any  $s'_0 < s_0 $. The  $b_{01}$-term in Eq.~\eqref{eq:bceff} is scale-dependent arising from the derivative of the window function. 
Typically, $W\to 1$ as $k\to 0$.  In this limit $b_{\rm W} (k) \to b_{10}$, so $b_{10}$ can be estimated reliably with little knowledge of the window.  However, the expression above shows that, to measure the bias parameter $b_{01}$ associated with scale dependent bias, one must model $W$ accurately.  
%Eq.~\eqref{eq:bceff} is similar to that in \cite{MussoParanjapeSheth2012} except we have adopted the effective window $W_{\rm eff}$.  There is a more refined excursion set peak model \cite{ParanjapeShethDesjacques2013}, in which a further condition that the point is a peak is imposed.  The first crossing constraint is more fundamental as it solves the cloud-in-cloud problem and gives the correct mass function in the excursion set theory. The additional term due to the curvature constraint is highly degenerate with the $b_{01}$-term  for the cross power spectrum. Thus only including the $b_{01}$-term captures most of the scale dependence already. We shall discuss more on the excursion set peak model in \cite{LagBiasPaper}, while we will focus on the model Eq.~\eqref{eq:bceff} here. 

A Gaussian smoothing window has long been a popular choice \cite[e.g.][]{BBKS86, EliaLudlowPorciani2012, BaldaufDesjacquesSeljak2014}.  However, the obvious oscillations at large $kR_{\rm Lag}$ in Fig.~\ref{fig:bc_TH_Eff_compare} rule this out as a viable model.  In contrast, the bias associated with setting $W=W_{\rm TH}$ in Eq.~\eqref{eq:bceff} will oscillate, though there is no guarantee that the phase or amplitude of the oscillations will match the measurements.  

The blue curves show the result of setting $W=W_{\rm TH}$ and treating $b_{10}$, $b_{ 01}$ and $R$ as free parameters when fitting $b_{\rm W}$ to our measurements.  (Because the high-$k$-part generally has smaller error bars compared to the low-$k$-part, to make sure the low-$k$-part is properly fitted, we first determine $b_{10}$ by fitting to $kR_{\rm Lag } \le 0.15$.  We then keep this value of $b_{10}$ fixed and fit $b_{01} $ and $R$ over the range $kR_{\rm Lag }\le 4$.  We do so because although the numerical uncertainty of the low-$k$-part is relatively large, its theoretical uncertainty is small.)  Evidently, the best-fit model over-predicts the amplitude of the oscillations.  This suggests that $W_{\rm eff}$ of Eq.~\eqref{eq:Weff}, a Gaussian smoothed tophat, may be able to provide a better description:  the Gaussian term will damp the oscillations.

The red curves show the result of replacing $W_{\rm TH}$ with $W_{\rm eff}$ in Eq.~\eqref{eq:bceff}.  Although this does provide a better description of the measurements, the amplitude of the oscillations in the best-fit model is still too large, especially at low masses.  Green curves show the result of allowing $f$ -- the parameter which controls the strength of the Gaussian damping -- to vary (we also extend the range over which we fit to $kR_{\rm Lag }\le 8$).  Now the fit is good for all masses, with a hint that the model ceases to provide a good description when $\nu_{\rm sc} \lesssim 1$.  Comparison of the curves shows that $W_{\rm TH}$ provides a good fit up to the peak in $b_c$, but it fails at larger $kR_{\rm Lag}$; on these scales, $W_{\rm eff}$ is much more accurate if $f$ is allowed to vary.  

Fig.~\ref{fig:bc_Lag_Dextrap} shows that Eq.~\eqref{eq:bceff} with $W_{\rm eff}$ and $f=1$ fits our measurements well also for the protohalos of halos identified at $z\sim 1$, although, as for $z=0$ halos, its performance starts to deteriorate when $\nu_{\rm sc} \lesssim 2 $.  Since the main focus of this paper is $W_{\rm eff}$, rather than the values of the best-fitting parameters themselves, we only provide these values in Appendix~\ref{sec:More_on_bc_fit}.  In~\cite{LagBiasPaper} we show how to use the best-fitting bias parameters, $b_{10}$ and $b_{01}$, to extract information about the physics of halo formation.

\section{ Excursion set peak mass function using the effective window}
\label{sec:nesp}

The original peak model \cite{BBKS86} considered peaks in the overdensity field identified on a fixed smoothing scale.  The excursion set approach \cite{BCEK1991} asserts that, to define protohalos, one must consider multiple smoothing scales.  While this multi-scale problem is difficult to handle analytically, the simpler problem, in which one considers one smoothing scale and the derivative on that same scale, is tractable and rather accurate \cite{MussoSheth2012}.  Refs.~\cite{AppelJones1990,ParanjapeSheth2012, ParanjapeShethDesjacques2013} showed how to merge the two approaches to define the excursion set peak (ESP) model.

The spectral moments $s_j$ play an important role in the ESP approach; Eq.~(\ref{eq:sj}) shows that they depend on the shape of the smoothing window.  The first derivations of the excursion set peak (ESP) mass function \cite{AppelJones1990, ParanjapeSheth2012} used Gaussian smoothing windows for all $s_j$.  In contrast, \cite{ParanjapeShethDesjacques2013} used different smoothing windows for different spectral moments -- in effect, they used tophat smoothing for the overdensity and its scale dependence (the excursion set constraint) and Gaussian smoothing for spatial dependence (the peak curvature constraint).  They also made a well-motivated approximation which derived from the fact that the peak curvature and excursion set variables are tightly correlated.  \cite{BiagettiChanetal_2014} showed that the approximation in \cite{ParanjapeShethDesjacques2013} differed from the exact expression by only a few percent; however, they too used different smoothing filters for the different $s_j$.  Since our $W_{\rm eff}$ leads to convergent $s_j$ for all $j$ of interest, it is natural to ask:  What is the associated ESP mass function?  It turns out that using the same smoothing window for all quantities which matter yields an expression which is essentially the same as Eq.~(10) of \cite{BiagettiChanetal_2014}, as we describe shortly.

First, we must account for the fact that the ESP predictions also depend on the critical overdensity required for a protohalo to collapse and form a halo.  This critical density is expected to vary from one halo to another, with a mean which is close to that associated with the spherical collapse model.  However, as noted by \cite{LagBiasPaper}, the precise value of the mean, and the scatter around it, will both depend on choice of filter.  \cite{ParanjapeShethDesjacques2013,BiagettiChanetal_2014} used a tophat filter when smoothing the overdensity field, since this is the filter which is usually assumed in the spherical collapse calculation.  Since we use $W_{\rm eff}$ throughout, we must check if the distribution of protohalo overdensities associated with $W_{\rm eff}$ differs from that in previous work. 

\begin{figure}[]
\centering
\includegraphics[width=\linewidth]{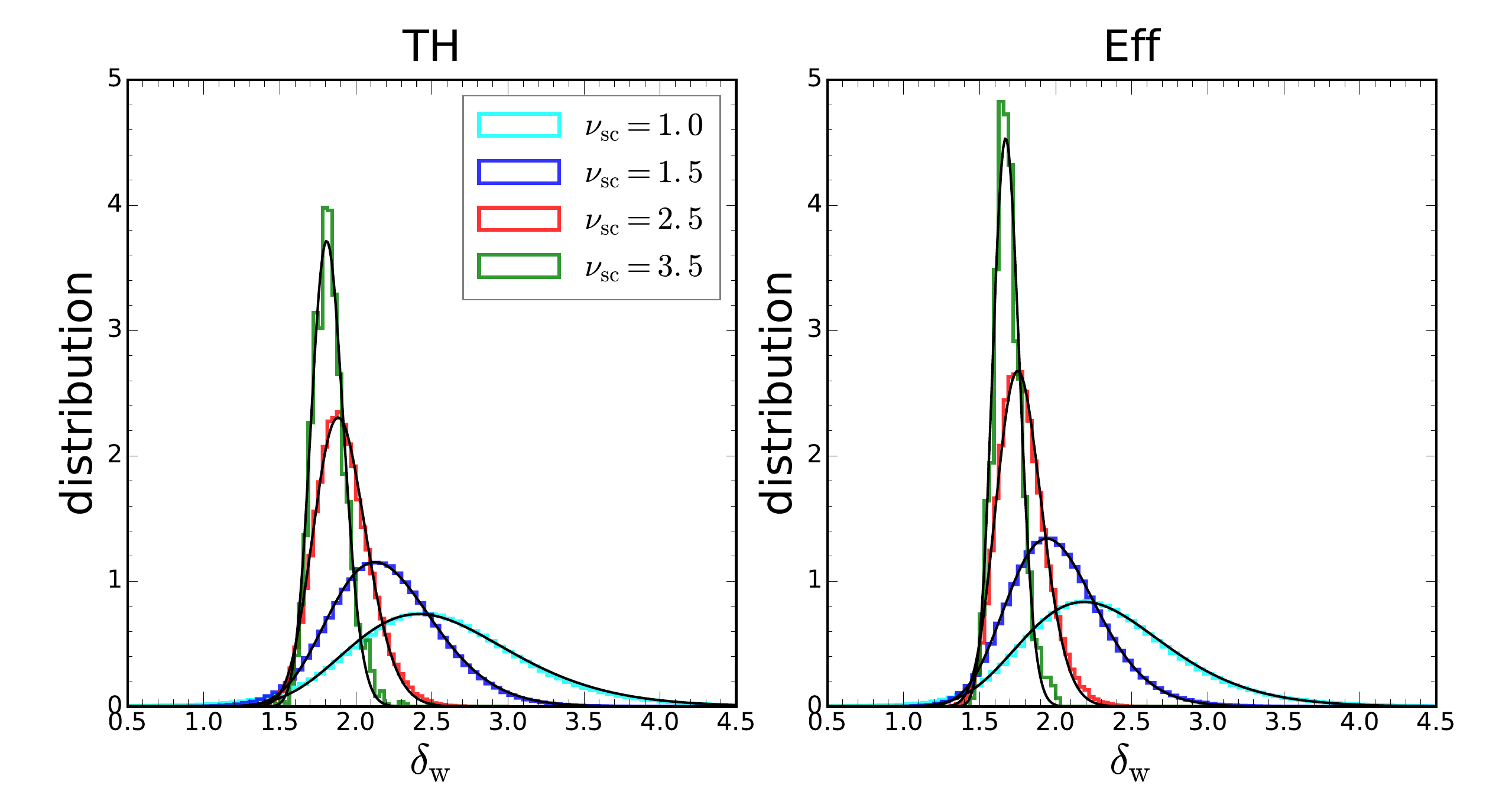}
\caption{Distribution of overdensity within protohalo patches for $\nu_{\rm sc} = 1$, 1.5, 2.5, and 3.5 (histograms), and the corresponding best-fit lognormal distributions (solid black line). Left and right panels show results for $W_{\rm TH}$ and $W_{\rm eff}$, respectively.  }
\label{fig:scatter_dist_lognormal_fit}
\end{figure}

Fig.~\ref{fig:scatter_dist_lognormal_fit} shows that there is indeed a difference.  The panels on the left and right show the mean overdensity within $W_{\rm TH}$ and $W_{\rm eff}$ for a range of halo masses, here parametrized by $\nu_{\rm sc} $, for which $\sigma_0$ is smoothed by $W_{\rm TH}$ or $W_{\rm eff}$ respectively.  Comparison of the histograms in the two panels shows that the mean and variance are slightly smaller for $W_{\rm eff}$:  $\langle\delta_{\rm TH}\rangle =  1.51 + 0.63 \sigma_0$ and $\langle\delta_{\rm eff}\rangle =1.42 + 0.55 \sigma_0$, with variance $\Sigma^2_{\rm TH} =  0.0861 \sigma_0^{2.66} $ and $\Sigma^2_{\rm eff} = 0.0645 \sigma_0^{2.67} $.  The difference in the mean is easily understood: the enclosed density is typically a decreasing function of scale, and $W_{\rm eff}$ is more extended than $W_{\rm TH}$.  The solid curves show the best-fitting lognormal distributions.  As we describe below, we use these lognormal fits to the measured distributions when generating our ESP mass functions.  

\subsection{Halo abundances}
The comoving number density of ESP peaks of mass $M$ is 
\beq
\label{eq:Mfn_formula}
 n(M) \equiv  \frac{ \rho_{\rm m}  }{ M  } \frac{ d \ln \nu_{\rm sc}  }{ d \ln M }\,
 \nu_{\rm sc} f_{\rm esp}(\nu_{\rm sc} ).
\eeq
In  $\nu_{\rm sc}  $,  $\sigma_0(R_{\rm Lag})$ is related monotonically to mass $M\equiv \rho_{\rm m}\,4\pi R_{\rm Lag}^3/3$, but this relation depends on the shape of the smoothing filter.  

The ESP multiplicity function $f_{\rm esp}$ is obtained by averaging its value $f_{\rm esp}( \nu_{\rm sc}|\epsilon)$ for a fixed `barrier' of  `height'
\beq
 \label{eq:sq_barrier}
 B(s_0) = \delta_0 + \epsilon \sqrt{ s_0 },
\eeq
over the distribution of heights, here parametrized by a distribution of $\epsilon$:  
\beq
\label{eq:fESP}
 f_{\rm esp}( \nu_{\rm sc} )
 = \int d\epsilon\, f_{\rm esp}( \nu_{\rm sc}|\epsilon)\, p(\epsilon).
 \eeq
 I.e., $B$ is the quantity we called $\delta_W$ in Fig.~\ref{fig:scatter_dist_lognormal_fit}, and its distribution is attributed to $\epsilon$ (rather than $\delta_0$).  Note that our $\epsilon$ is $\beta$ in \cite{ParanjapeShethDesjacques2013}, and our $f_{\rm esp}( \nu_{\rm sc}|\epsilon)$ is
\beqa
 f_{\rm esp}(\nu_{\rm sc}|\epsilon) &=& \frac{ M/ (\bar\rho\,V_* )}{\gamma_{\nu u}\nu_{\rm sc}}
 \frac{e^{-\nu^2/2}}{\sqrt{2\pi}} \nn\\
 &&\times \int_0^\infty dx\,f(x)\,\frac{e^{-(x-\gamma_{x\nu}\nu)^2/ [2(1-\gamma^2_{x\nu})] }}{\sqrt{2\pi(1-\gamma^2_{x\nu})}} \nn \\
 && \times \int_{u_0}^\infty du\, (u- u_0 ) \, p(u|\nu,x),
\eeqa
where
 $\nu = B/\sigma_0 = (\delta_0/\delta_{\rm sc})\, \nu_{\rm sc} + \epsilon$ is the peak height variable in units of its rms on scale $R_{\rm Lag}$, 
 $x\equiv -\nabla^2\delta/\sqrt{s_2}$ is the peak curvature variable in units of its rms on scale $R_{\rm Lag}$,
 $u_0\equiv  (dB/ds_0)/\sigma_u $ is the excursion set variable in units of its rms $\sigma_u \equiv \langle ( d\delta/ds_0)^2\rangle^{1/2}$, evaluated at $s_0(R_{\rm Lag}) $,
 $V_* \equiv (6\pi\,s_1/s_2)^{3/2}$,
 and $f(x)$ is given by Eq.(A15) of \cite{BBKS86}.
%%  \beqa
%%  f_{\rm esp}(\nu_{\rm sc}|\epsilon) &=& \frac{ V_{\rm TH} /V_*}{\gamma_{\nu u}\nu_{\rm sc}}
%%  \frac{e^{-\nu^2/2}}{\sqrt{2\pi}} \nn\\
%%  &&\times \int_0^\infty dx\,f(x)\,\frac{e^{-(x-\gamma_{x\nu}\nu)^2/2(1-\gamma^2_{x\nu})}}{\sqrt{2\pi(1-\gamma^2_{x\nu})}} \nn \\
%%  && \times \int_{B'}^\infty du\,\frac{u-B'}{\sigma_u}\,p(u|\nu,x),
%% \eeqa
%% where
%%  $  B' = dB/ds_0$,
%%  $u \equiv d\delta/ds_0$, $\sigma_u^2 \equiv \langle u^2\rangle$, 
%%  $\gamma_{\nu u} = (2\sigma_u\sigma_0)^{-1}$ is the normalized cross-correlation coefficient between $\delta$ and $u$,
%%  $x \equiv -\nabla^2\delta/\sqrt{s_2}$,
%%  $\nu = B/\sigma_0 = (\delta_0/\delta_c)\, \nu_{\rm sc} + \epsilon$, 
%%  $V_* \equiv (6\pi\,s_1/s_2)^{3/2}$,
Here $\gamma_{\nu u} = (2\sigma_u\sigma_0)^{-1}$  and  $\gamma_{x \nu} = s_1 / (\sigma_0 \sigma_2 )$ are the normalized cross-correlation coefficients between $\delta$ and $u$, and $\delta$ and $x$, respectively.   The distribution $p(u | \nu,x )$ is a conditional Gaussian distribution, so the integral over $u$ yields
\beq
 \bar u\, \left[\frac{1 + {\rm erf}(\bar u/ ( \sqrt{2}\Sigma ) )}{2}
 + \frac{\Sigma}{\bar u} \,\frac{{\rm e}^{-\bar u^2/ (2\Sigma^2 )}}{\sqrt{2\pi}}\right],
\eeq
where 
\beq
 \bar u \equiv x\,\frac{\gamma_{xu} - \gamma_{x\nu}\gamma_{\nu u}}{1 - \gamma_{\nu x}^2} + \nu\,\frac{\gamma_{\nu u} - \gamma_{\nu x}\gamma_{xu}}{1 - \gamma_{\nu x}^2} - u_0 ,
\eeq
and
\beq
 \Sigma^2 = \frac{1 - \gamma_{\nu x}^2 - \gamma_{\nu u}^2 - \gamma_{xu}^2 + 2\gamma_{\nu x}\gamma_{\nu u}\gamma_{xu}}{1 - \gamma_{\nu x}^2}.
 \eeq
(As our notation suggests, $ \gamma_{xu} $ denotes the cross correlation coefficient between $x$ and $u$).

Our Eq.~(\ref{eq:fESP}) is the same as Eq.~(10) of \cite{BiagettiChanetal_2014}, except that our $\epsilon$ and $x$ are their $\beta$ and $u$, our $u$ is proportional to their $-\mu$ (to conform with more standard notation), we are more careful about the limits of integration on our $u$ (this gives the extra $u_0$ term in $\bar u$), our $\Sigma^2$ corrects what appears to be a typographical error in their Eq.~(14), and we use the same smoothing window, $W_{\rm eff}$, in all the expressions for covariances between variables (e.g., their Eqs.~3--5 and~10).  Eq.~(14) of \cite{ParanjapeShethDesjacques2013} approximates $p(u|\nu,x)$ as a delta function centered on $x$ (for Gaussian smoothing $\gamma_{xu}=1$ and $\gamma_{\nu x} = \gamma_{\nu u}$), and uses Tophat smoothing for $\delta$ but Gaussian smoothing for $x$, with $R_{\rm G}\approx R_{\rm TH}/\sqrt{5}$ (which comes from matching $W_{\rm TH}(x)$ and $W_{\rm G}(x)$ to order $x^2$).

\begin{figure}[]
\centering
\includegraphics[width=\linewidth]{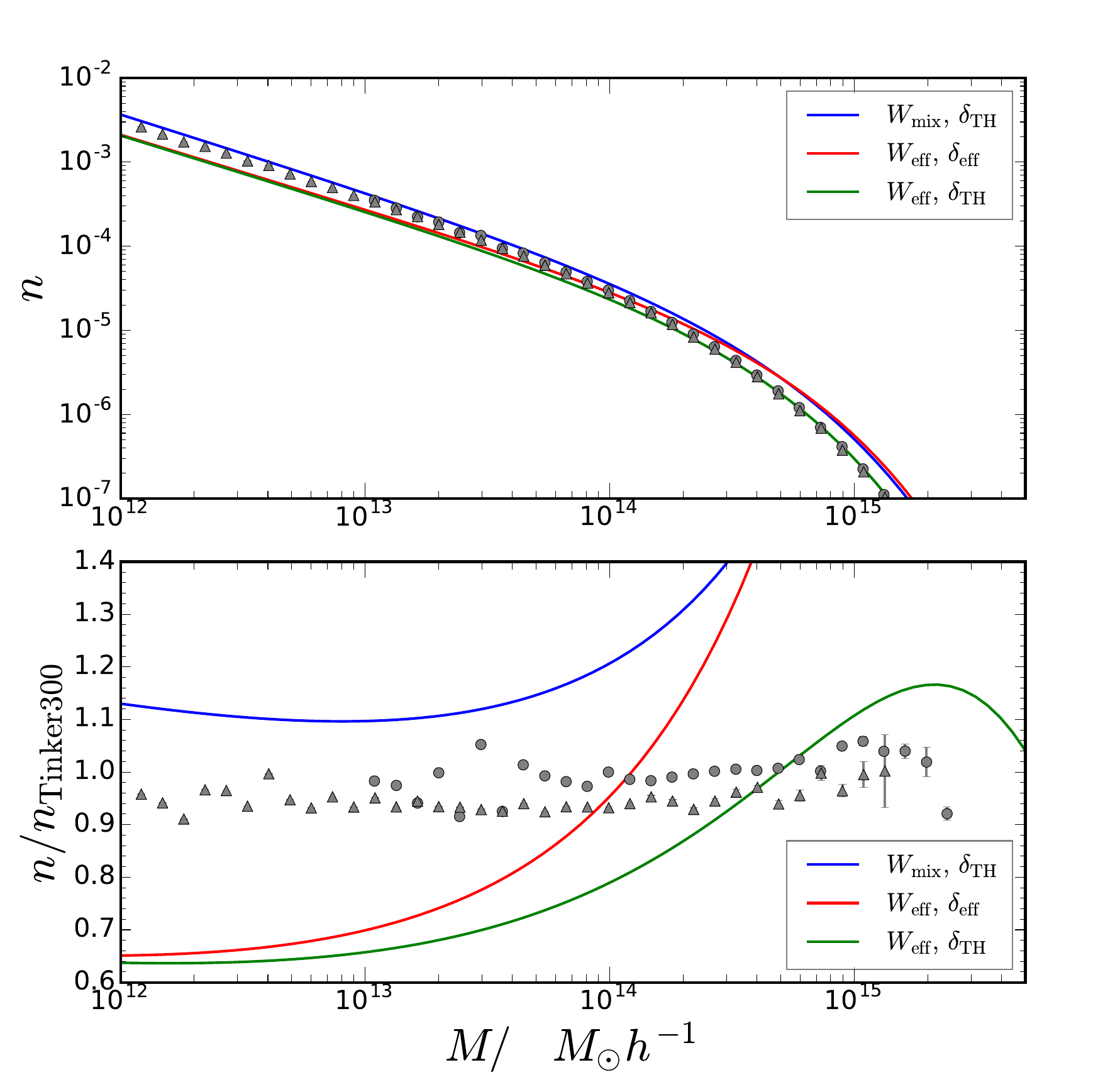}
\caption{Top: The $z=0$ mass function $n$ measured in simulations (triangles for Carmen, circles for Oriana). Curves show three ESP predictions as labeled in the legend.  Bottom:  Mass functions normalized by the fitting formula from Ref.~\cite{2008ApJ...688..709T} for halos having spherical overdensity $\Delta =300$. }
\label{fig:mfn_esp}
\end{figure}

The symbols in Fig.~\ref{fig:mfn_esp} show the $z=0$ mass function measured in the simulations:  triangles and circles represent Carmen and Oriana.  The curves in the top panel show the ESP prediction, Eq.~\eqref{eq:fESP}, with three different pairings of the smoothing window and the critical overdensity.  The solid blue curve uses $W_{\rm TH}$ for $\delta$ but $W_{\rm G}$ for $x$, so we loosely refer to it as having `mixed' smoothing, $W_{\rm mix}$.  Since $\delta$ is smoothed with a tophat, we use our fits to the distributions shown in the left hand panel of Fig~\ref{fig:scatter_dist_lognormal_fit} for the associated critical overdensity.  The red curve uses $W_{\rm eff}$, so the critical overdensity comes from fitting to the results shown in the right hand panel of Fig~\ref{fig:scatter_dist_lognormal_fit}.  The green curve uses $W_{\rm eff}$ but $\delta_{\rm TH}$ as a crude way of accounting for the fact that $W_{\rm eff}$ does not select a random subset of particles, but the subset with larger infall speeds (c.f. Appendix~\ref{sec:sc}).

To reduce the dynamic range, the lower panel shows the mass function normalized by the fitting formula of Ref.~\cite{2008ApJ...688..709T} for the abundance of spherical overdensity halos identified because they are $\Delta =300$  times denser than the background.  This formula is similar to that of Ref.~\cite{2016MNRAS...456...2486}.  This format makes it easy to see that the $W_{\rm mix}$ prescription works reasonably well at smaller masses, though the agreement is not quite as good as demonstrated by Ref.~\cite{ParanjapeShethDesjacques2013} for $\Delta=200$ SO halos.  However, it overpredicts the counts at larger masses.  Our $W_{\rm eff}$ also overpredicts at large masses, but underpredicts by nearly a factor of 2 at smaller masses.  Using $\delta_{\rm eff}\to\delta_{\rm TH}$ with $W_{\rm eff}$ produces good agreement at large masses, but does not fix the problem at smaller masses.

\subsection{Lagrangian protohalo bias}
We remarked in the Introduction that there is a tight connection between halo abundances and clustering.  So, having compared ESP mass functions for $W_{\rm eff}$ and $W_{\rm mix}$, we now discuss their associated ESP bias factors.

In the ESP approach, protohalo patches satisfy three constraints:  their enclosed density must be large enough; this density must be larger than that on the next larger smoothing scale (the ES in ESP); and it must be a local peak (the P in ESP).  Therefore, $b_c$ is expected to be the sum of three terms, each having slightly different $k$-dependence:  the first scales as $W$, the second as $dW/d\ln s_0$, and the third as $k^2 W$.  For a Gaussian smoothing filter, the latter two are exactly degenerate.  %which is why only $\nu$ and $x$ appear in Eq.~\eqref{eq:fESP_e_Gaussian} for $f^{\rm G}_{\rm ESP}$.
For more general filters, it would be reasonable to expect that one can model $b_c$ quite accurately by using $W$ and either $dW/d\ln s_0$ or $(kR)^2W$, but it may not be necessary to use both \cite{LagBiasPaper}.  The model fits in Figs.~\ref{fig:bc_TH_Eff_compare} and~\ref{fig:bc_Lag_Dextrap} show that this indeed seems to be the case for $W_{\rm eff}$; Eq.~\eqref{eq:bceff}, which does not have a $(kR)^2W$ term, provides a good description of the measurements.

However, this is not as straightforward for $W_{\rm mix}$, since the approach uses $W_{\rm TH}$, $dW_{\rm TH}/d\ln s_0^{\rm TH}$ and $(kR_{\rm G})^2 W_{\rm G}$ with $R_{\rm G}=R_{\rm TH}/\sqrt{5}$.  In this case, the $kR\gg 1$ dependence of $W_{\rm TH}$ is very different from that of $W_{\rm G}$.  The blue curves in Fig.~\ref{fig:bc_TH_Eff_compare} show that Eq.~\eqref{eq:bceff} with $W_{\rm TH}$ is a bad choice; its oscillations at large $kR_{\rm TH}$ are too strong.  In addition, that there are oscillations at all mean that using $W_{\rm G}$ in Eq.~\eqref{eq:bceff} will not work either.  Finally, approximating
\beq
\label{eq:bcmix} 
 b_{\rm mix}(k) = b^{\rm mix}_{10}\,W_{\rm TH}(kR_{\rm TH}) 
 + b^{\rm mix}_{01}\,\frac{s^{\rm TH}_0}{s^{\rm mix}_1}\,k^2\, W_{\rm G}\left(\frac{kR_{\rm TH}}{\sqrt{5}}\right),
 \eeq
 where $s_1^{\rm mix}$ is given by Eq.~\eqref{eq:sj}, but with $W^2\to W_{\rm TH}W_{\rm G}$ is not accurate either; becuuse there is no oscillations in $W_{\rm G}$, the wiggles from the $b_{10}$-term is too weak compared to the numerical results.

Ref.~\cite{LagBiasPaper} suggest that one might define linear combinations of the $dW/d\ln s_0$ and $k^2W$ contributions -- a suitably weighted sum, $T_+$, and difference, $T_-$, of the two terms.  (The idea is that $T_-$ should only matter at larger $kR_{\rm Lag}$.)  Although we do not show it here, we have found that using only $W_{\rm TH}$ and $T^{\rm mix}_+$ still leads to unacceptably large oscillations at large $kR_{\rm Lag}$.  We conclude that although the $W_{\rm mix}$-based approach provides a slightly better description of halo abundances, our $W_{\rm eff}$-approach provides a more efficient description of Lagrangian protohalo bias.

\section{ Conclusions} 
\label{sec:conclusion}
It is often assumed that protohalos in the initial conditions can be defined using a spherical tophat window function. This is, of course an idealization.  Other than being a convenient approximation in the context of the spherical collapse model, there is no fundamental reason why the protohalo profile should be a spherical tophat. Realistically, we expect some particles within the Lagrangian protohalo patch -- those with insufficiently large infall speeds -- will not be incorporated into the final object, while others which were initially more distant may be, provided they fell in quickly enough.  Thus we expect that, if the protohalo particles are those whose infall times are smaller than the present time, then the spatial distribution of the spherically averaged protohalo profile will be more extended than a tophat (Appendix~\ref{sec:sc}).

We explored this issue by measuring the Lagrangian protohalo profile in numerical simulations.  We did so by identifying halos of similar masses, stacking their protohalo patches together, and measuring the spherically averaged profile of the stack.  We found that the Lagrangian protohalo profile is indeed more extended than a tophat, but less extended than a Gaussian (Fig.~\ref{fig:nh_ntot_ratio}).  
%In particular, on average there is still a few per cents of the total particles near the center of mass of the Lagrangian halo  do not make it into the final Eulerian halos. 
In Fourier space, the profile $W_{\rm eff}$ is well approximated by the product of a tophat and a Gaussian (Eq.~\eqref{eq:Weff}).  The convolution which describes the real space profile can be done analytically (Eq.~\eqref{eq:WG*THrealspace}) so most analyses of halo abundances and clustering, which assume tophat or Gaussian profiles, can be performed with little modification.

For example, in excursion set approaches, the protohalo-matter cross power spectrum is the product of a scale-dependent bias factor and the window function (Eq.~\eqref{eq:bceff}).  Since the two effects appear combined in any measurement, to measure the scale-dependence of bias, the window function must be modeled accurately.  We find that using $W_{\rm eff}$ of Eq.~\eqref{eq:Weff} in Eq.~\eqref{eq:bceff} describes the measured cross spectrum well up to $kR_{\rm Lag} \sim 10 $ (Figs.~\ref{fig:bc_TH_Eff_compare} and~\ref{fig:bc_Lag_Dextrap}).  Importantly, the same window function also describes the protohalo profile in real space (Fig.~\ref{fig:nh_ntot_ratio}).  %This justifies our assertion that, to a good approximation, $W_{\rm eff}$ can be thought of as playing two different roles (see Appendix~\ref{sec:sc}).  

In the excursion set peak approach, predicting the abundance and clustering of halos requires the evaluation of the spectral moments in Eq.~(\ref{eq:sj}).  Some of these integrals diverge for a tophat filter, and this complicates how the approach is implemented.  In contrast, they are all well defined for our $W_{\rm eff}$, so this allows a more straightforward ESP analysis (Sec.~\ref{sec:nesp}).  The predicted mass functions also depend on the critical threshold overdensity required for collapse.  We use the values measured in the simulations (Fig.~\ref{fig:scatter_dist_lognormal_fit}), leaving no free parameters. The predicted ESP mass function, Eq.~\eqref{eq:fESP}, is not more accurate than the old mixed-filtering prediction; it significantly underpredicts the counts at lower masses (Fig.~\ref{fig:mfn_esp}).  On the other hand, using the critical overdensity associated with tophat smoothing results in better agreement, especially at high masses.  This suggests that a more fully Lagrangian model, built on infall speeds rather than smoothed overdensity, may be more realistic (Appendix~\ref{sec:fuzzy}) and ultimately more accurate.

On the other hand, our $W_{\rm eff}$ approach does provide a better description of the scale dependence of bias, which it describes to $kR_{\rm Lag}\gg 1$ for halos with $\nu_{\rm sc}\ge 1$.  The accuracy of our estimated bias parameters allows us to check the consistency relations that these bias parameters should satisfy \cite{MussoParanjapeSheth2012, ParanjapeShethDesjacques2013}.  In \cite{LagBiasPaper}, we show that this accuracy allows us to extract information about physics of halo formation from measurements of the large-scale bias.

\section*{Acknowledgments} 
We thank A. Paranjape for insightful discussions, V. Desjacques for comments on an early draft of the paper, and the anonymous referee for constructive comments that improved the paper.  We acknowledge the LasDamas project \footnote{\url{http://lss.phy.vanderbilt.edu/lasdamas}} for the simulations used in this work, which were run using a Teragrid allocation as well as other RPI and NYU computing resources.  KCC thanks the ICTP in Trieste and the theory group of CERN, where part of the work was done, for their hospitality, and acknowledges support from the Swiss National Science Foundation and the Spanish Ministerio de Economia y Competitividad grant  ESP2013-48274-C3-1-P.

\appendix

\section{Systematic effects}\label{sec:sys}

\subsection{Discreteness of the Lagrangian grid}\label{sec:discrete}
\begin{figure}
\centering
\includegraphics[width=\linewidth]{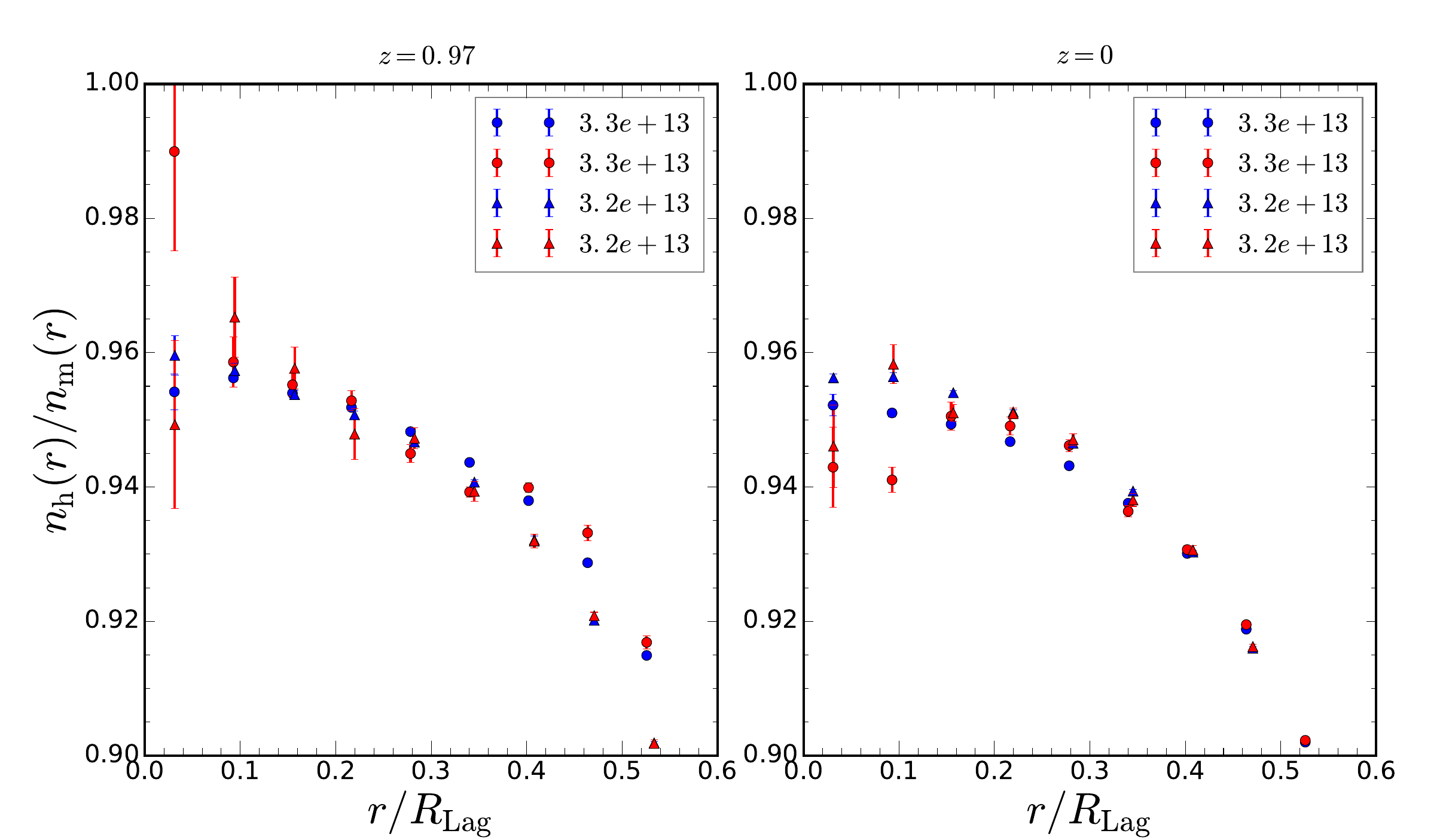}
\caption{Comparison of $p_{\rm h}=n_{\rm h}/n_{\rm m}$ values for protohalos in the same narrow mass bin (legend gives masses in units of $h^{-1}M_\odot$) in the Carmen (triangles) and Oriana (circles) simulations.  Each Carmen halo has about 670 particles, and this is about $10\times$ that in each Oriana halo.  Red symbols use the mean particle density $\bar n_{\rm m}$ for the denominator $n_{\rm m}$; blue symbols show when both $n_{\rm h}$ and $n_{\rm m}$ are measured in the simulations.  At small $r/R_{\rm Lag}\le 0.2$, the red circles indicate smaller $p_{\rm h}$ than the blue circles, which themselves lie below the triangles: on these scales the discreteness of the grid has compromised the estimate of $p_{\rm h}$ for the low resolution protohalo.  
 }
\label{fig:discrete}
\end{figure}

Fig.~\ref{fig:discrete} compares $p_{\rm h} = n_{\rm h}/n_{\rm m}$ values for halos in the same narrow mass bin in the Carmen (triangles, about 670 particles) and Oriana (circles, about 66 particles) simulations. The blue symbols estimate both $n_{\rm h}$ and $n_{\rm m}$ from the particle distribution in the simulations, as we do in the main text.  The red symbols use the mean particle density $\bar n_{\rm m}$ for the denominator $n_{\rm m}$. This continuous approximation is noisy, especially for the first bin shown.  The red circles drop dramatically at small $r/R_{\rm Lag}\le 0.2$; discreteness effects clearly compromise these estimates in the lower resolution halo.  On these scales, the blue circles lie well above the red circles, indicating that the estimator which uses the actual particles performs better. On the other hand, the blue circles lie slightly below the triangles:  even with this estimator, discreteness is still an issue for the low resolution simulation.  In contrast, the red and blue triangles are rather similar:  discreteness is not an issue for the higher resolution halo.  We conclude that some of the offsets of $p_{\rm h}$ from unity at small $r$ are systematics and not physical.  

\begin{figure}
\centering
\includegraphics[width=\linewidth]{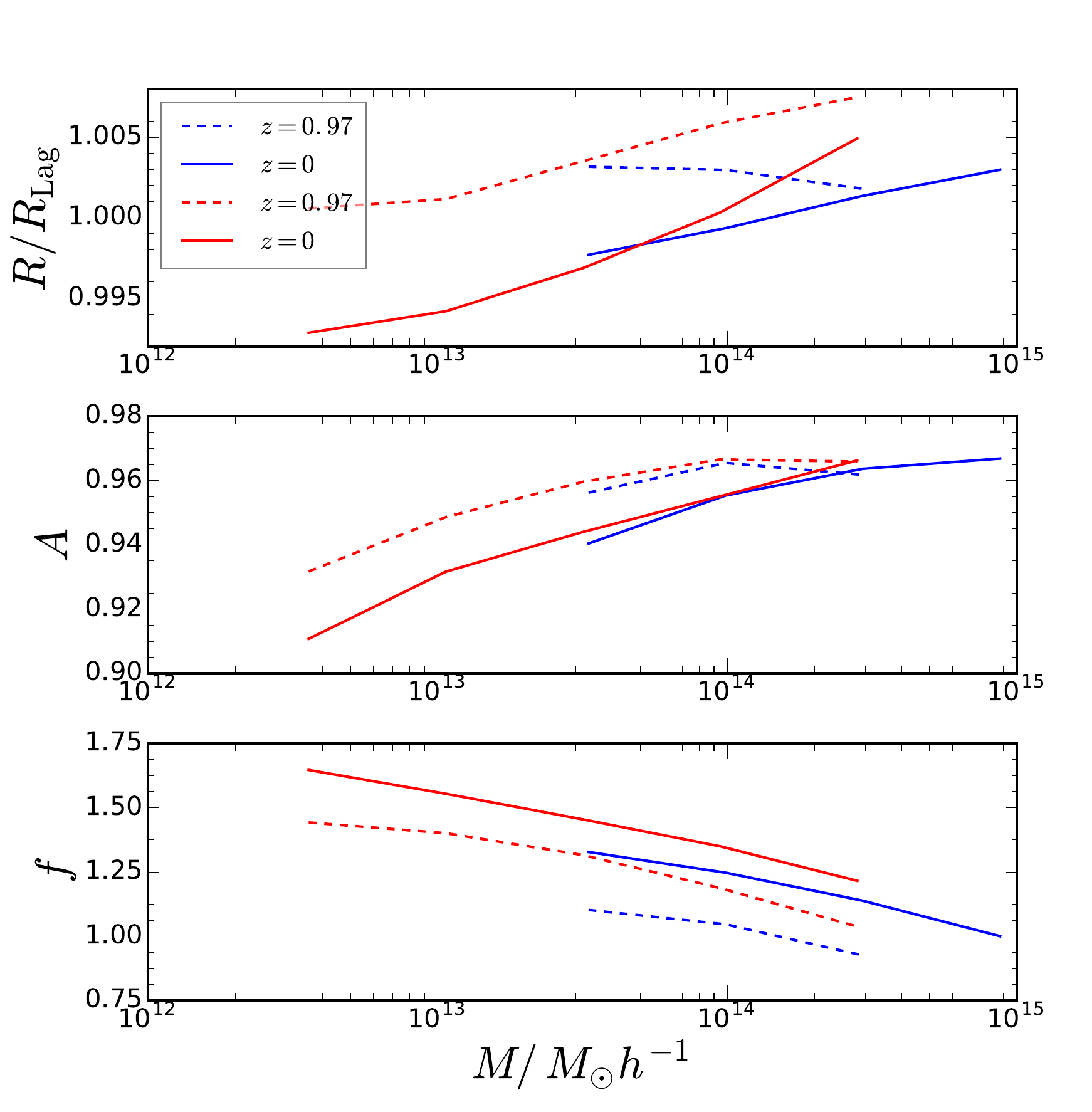}
\caption{ Same as Fig.~\ref{fig:LagrProfile_bestfit_WGWTH}, except that the best-fit parameters $\rho$, $A$ and $f$ for the Lagrangian protohalos are now shown as a function of halo mass. Red and blue curves show results for $z=0.97$ (dashed) and $z=0$ (solid) protohalos in the high and low resolution simulations (Carmen and Oriana).    }
\label{fig:LagrProfile_bestfit_WGWTH_M}
\end{figure}

In the main text, we used the parameter $A$ to account for the fact that $p_{\rm h}$ did not asymptote to unity at small $r$.  If some of the departures from unity which we show in Fig.~\ref{fig:LagrProfile_bestfit_WGWTH} are systematics associated with discreteness, then we expect them to scale with protohalo mass.  Fig.~\ref{fig:LagrProfile_bestfit_WGWTH_M} shows those results as a function of $M$ rather than $\nu_{\rm sc}$.  At fixed $M$ and $z$, we do not see strong differences between the two simulation sets.  Moreover, the dependence on $z$ is reduced when shown as a function of $\nu_{\rm sc}$ (as in the main text).  Thus, Fig.~\ref{fig:LagrProfile_bestfit_WGWTH_M} suggests that most of the tendency for higher $\nu_{\rm sc}$ to have larger $A$ is real.  This may be related to the fact that small $\nu$ protohalos tend to be less spherical.  

\subsection{Estimating the Lagrangian window from the displaced field} 
\label{sec:displaced}
\begin{figure}
\centering
\includegraphics[width=\linewidth]{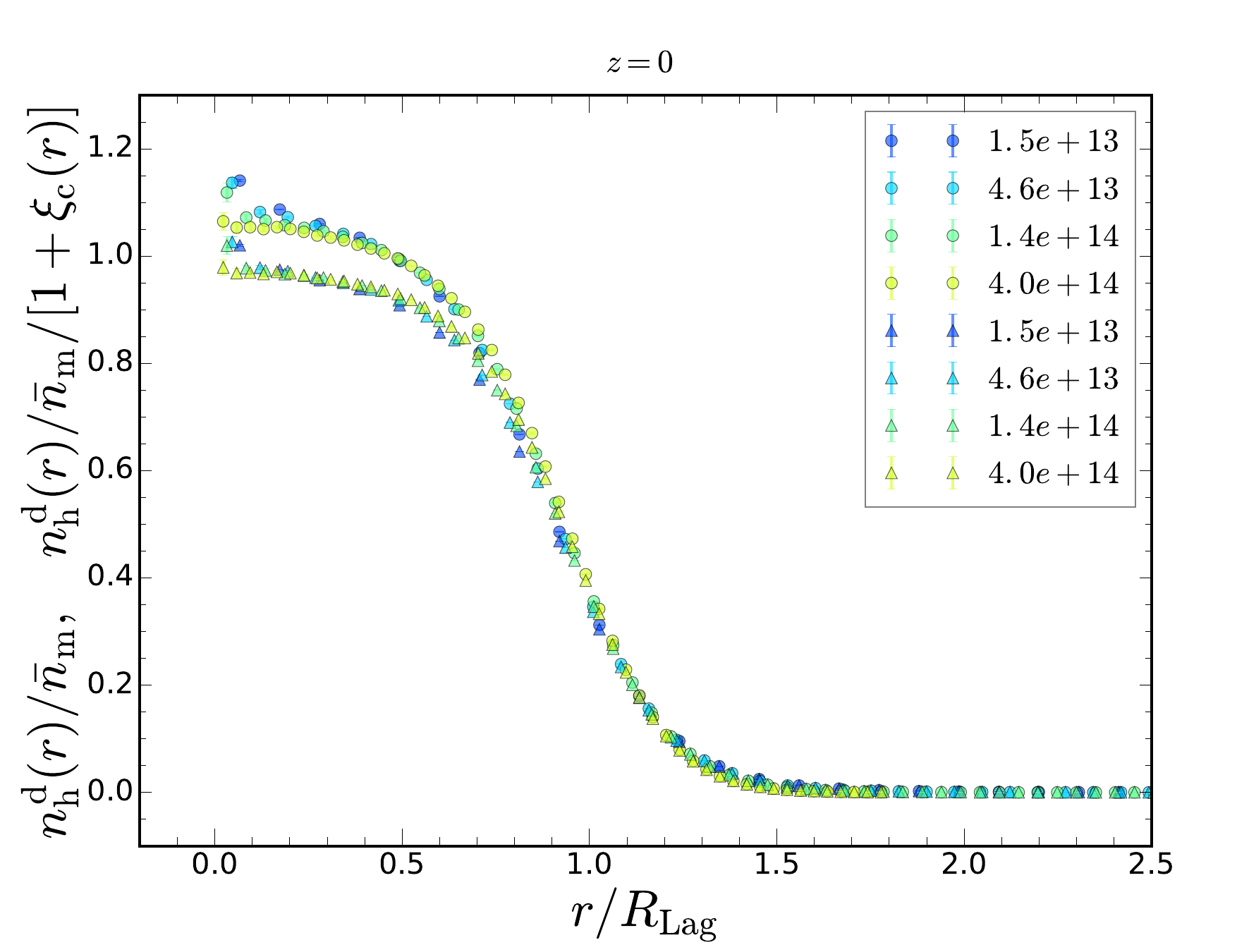}
\caption{  The ratio $ n_{\rm h}^{\rm d} / \bar{n}_{\rm m} $ obtained using displaced particles at the initial redshift $z_* =49$ (circles, the upper set of curves)  and when the clustering effect has been corrected $(n_{\rm h}/\bar{n}_{\rm m})/( 1 + \xi_{\rm c}) $ (triangles, the lower set).  Results are for the Lagrangian protohalos of the $z=0$ halos in the higher resolution (Carmen) simulations; legend gives the mean mass of each sample in units of $h^{-1}M_\odot$.}
\label{fig:rho_ratio_cluster_corr_z0_Carmen}
\end{figure}

We now discuss a method which uses the displaced particles to estimate $ p_{\rm h} $. As we mentioned in Sec.~\ref{sec:LagWin_real}, in this case we cannot replace $n_{\rm m}$ by $\bar{n}_{\rm m}$. Due to the presence of the halo at the origin, we must set 
\beq
 n_{\rm m}(r)  = \bar{n}_{\rm m} [ 1 + \xi_{\rm c} ( r ) ],  
\eeq
where $ \xi_{\rm c}$ is the cross correlation function between halo and matter. We can either measure $ \xi_{\rm c} $ directly or model it. To model $\xi_{\rm c}$ we use the bias model Eq.~\eqref{eq:bceff}  given  together with the bias parameters computed from the peak model \cite{ParanjapeShethDesjacques2013}  to obtain the cross power spectrum. Then  $\xi_{\rm c} $ follows after an inverse Fourier transform.  As in this paper, we focus on the window function, we will present the details of the model and the comparison with numerical measurements elsewhere \cite{LagBiasPaper}. We find that the model we adopted describes the simulation data well. Hence we will use $\xi_{\rm c} $ obtained from theory here. In Fig.~\ref{fig:rho_ratio_cluster_corr_z0_Carmen}, we compare the measurement  $n_{\rm h}^{\rm d} / \bar{n}_{\rm m} $ and    $(n_{\rm h}^{\rm d} / \bar{n}_{\rm m}) / ( 1 + \xi_{\rm c} )  $, where the superscript d in  $n_{\rm h}^{\rm d}$ emphasizes that it is estimated using the displaced particle distribution.  We have used the Lagrangian halos at $ z=0 $ from the Carmen simulation in this plot.  The clustering correction is mainly important  for $r \lesssim R_{\rm Lag} $, while for larger $r$,  $ \xi_{\rm c} \ll 1$. Due to the clustering enhancement,  $n_{\rm h}^{\rm d} / \bar{n}_{\rm m} $ exceeds 1 for  $r / R_{\rm Lag} \lesssim 0.5 $. After the correction, we find that the results agree with those from the method described in the main text very well (right panel of Fig.~\ref{fig:nh_ntot_ratio}).  This method is interesting in its own right:  That we are able to correct for the clustering effect is a non-trivial consistency check of our approach.

\begin{figure}
\centering
\includegraphics[width=\linewidth]{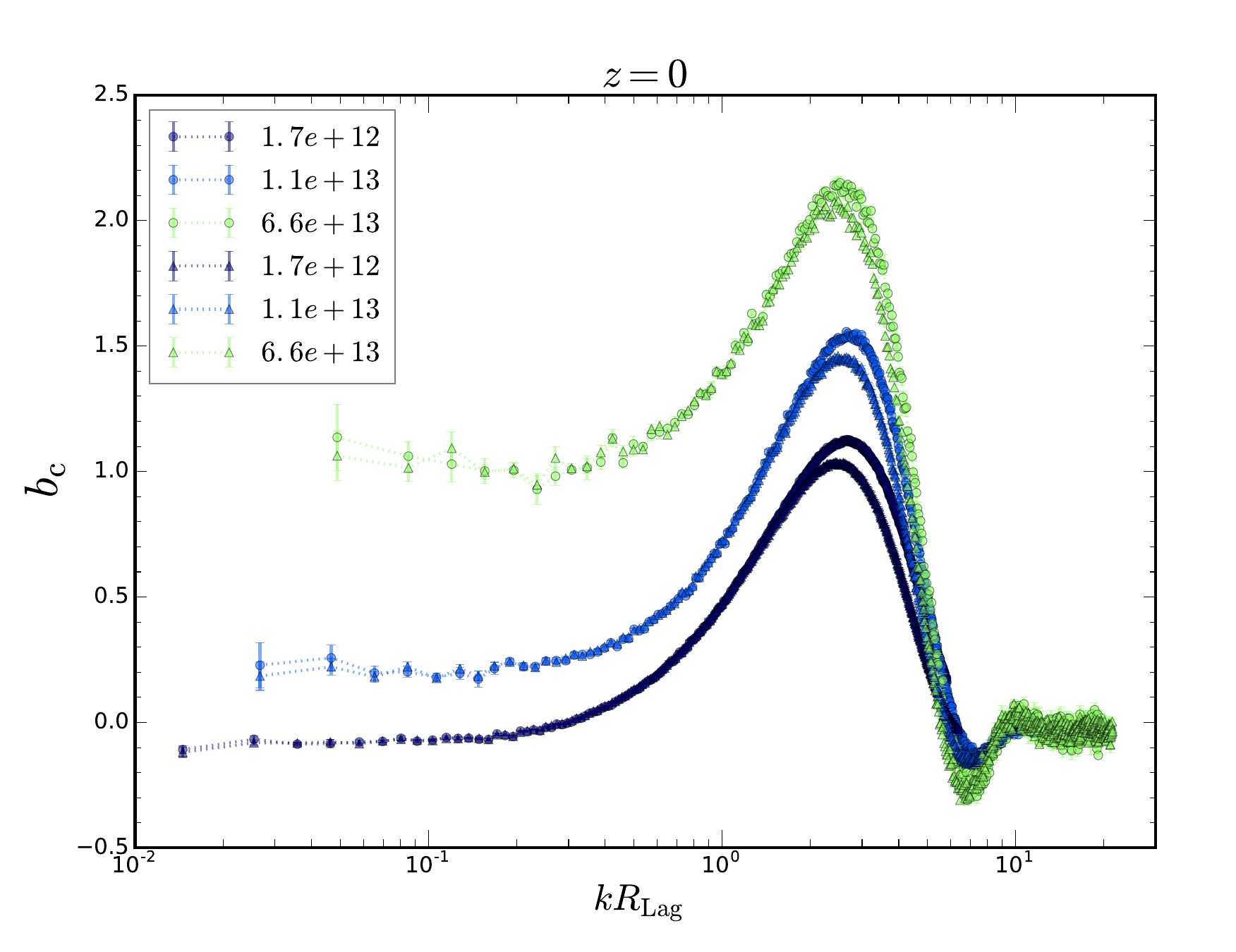}
\caption{ Lagrangian bias $b_{\rm c}$ obtained by cross correlating the Lagrangian halo field with the initial Gaussian random field (triangles) and the 2LPT particle distribution field (circles) for a range of protohalo masses (legend gives masses in units of $h^{-1}M_\odot$) for the $z=0$ halos in the higher resolution (Carmen) simulations. }
\label{fig:bc_DM_GRF_Evolve}
\end{figure}

The main text used the initial Gaussian random field to compute the cross power spectrum. It is worth pointing out how $b_{\rm c  }$ in Fourier space changes when the evolved dark matter field is used instead.  Fig.~\ref{fig:bc_DM_GRF_Evolve} compares $b_{\rm c}$ obtained using the initial Gaussian random field (triangles) with that for the 2LPT particle distribution (cirlces).  At low $k$, these two cases coincide as we do not expect this evolution to modify the $ b_{01}$-part although the one from the particle distribution is slightly noisier. Both cases show similar oscillatory patterns. However, deviations arise around $kR_{\rm Lag}\sim 2$, where $b_{\rm c}$ from the evolved field is slightly larger.  

Figs.~\ref{fig:rho_ratio_cluster_corr_z0_Carmen} and \ref{fig:bc_DM_GRF_Evolve} demonstrate that the evolution of the dark matter field does not change the large scale, but the small scale ($ r/R_{\rm Lag} \lesssim 1$ or $kR_{\rm Lag} \gtrsim 1$) is enhanced by evolution.

\subsection{ Best-fitting cross-bias parameters}
\label{sec:More_on_bc_fit}
The main text shows that $W_{\rm eff}$, together with the excursion set bias model, provides a better description of the protohalo-matter cross bias parameter than does a tophat window $W_{\rm TH}$.

\begin{figure*}
\centering
\includegraphics[width=\linewidth]{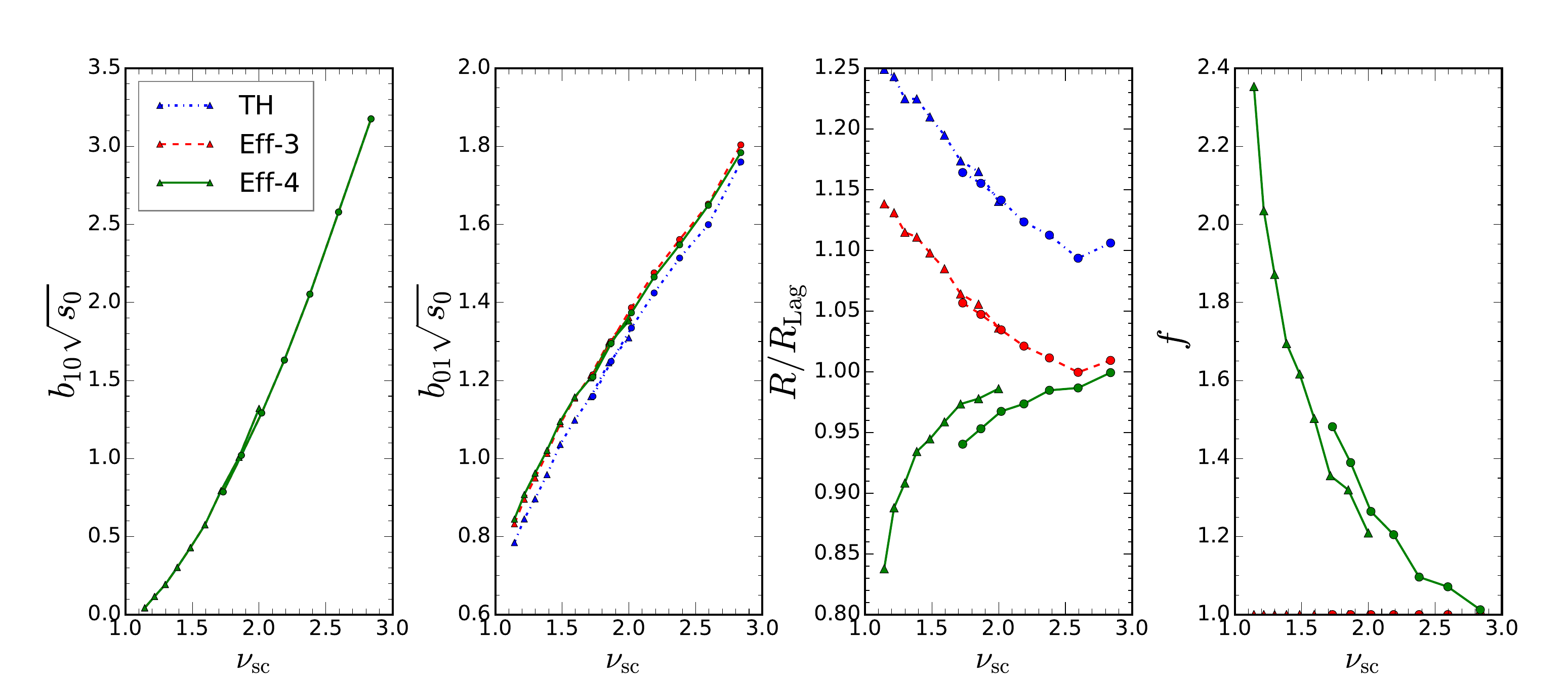}
\caption{ The best fit parameters $b_{10}$, $b_{01}$, $R$ and $f$ determined from fitting Eq.~\eqref{eq:bceff} to the protohalo-matter cross correlation measurements shown in Fig.~\ref{fig:bc_TH_Eff_compare}.  Dotted blue and dashed red lines show results associated with $W_{\rm TH}$ and $W_{\rm eff}$ when $f=1$; solid green uses $W_{\rm eff}$ and allows $f$ to also vary. The measurements are from Oriana (circles) and Carmen (triangles) at $z=0$.   }
\label{fig:bestfit_TH_Eff_compare}
\end{figure*}

Fig.~\ref{fig:bestfit_TH_Eff_compare} shows the best fit parameters for these fits.  When $f$ is set to unity, and only $b_{10}$, $b_{01}$ and $R$ are varied, then we only attempt to fit to the measurements at $kR_{\rm Lag} \le 4$; when $f$ is also varied, then we fit up to  $kR_{\rm Lag} \le 8$.  As we noted in the main text, $ b_{10} $ is insensitive to the window function adopted, and this is especially true for our fitting procedure. For $b_{01}$, the two $W_{\rm eff}$ fits (i.e. with $f=1$ and variable $f$) yield similar results, while the value associated with $W_{\rm TH}$ is slightly lower.

For $W_{\rm TH}$, the best-fit $R$ is larger than $R_{\rm Lag}$ by at least 10\%.  For $W_{\rm eff}$, $R$ agrees with $R_{\rm Lag}$ for $ \nu_{\rm sc} \gtrsim 2.5 $, but it too exceeds $R_{\rm Lag}$ as $\nu_{\rm sc} $ decreases.  This can be understood as follows.  As $\nu_{\rm sc} $ decreases, higher damping is required (e.g.~Fig.~\ref{fig:bc_TH_Eff_compare}). If $f$ is fixed to unity, then $R_{\rm G}=R_{\rm TH}/5$ so this damping can only be incorporated by increasing $R_{\rm TH}$. When $f$ is allowed to vary, we find it becomes substantially larger than unity, meaning $R_{\rm G}$ can increase so $R_{\rm TH}$ no longer needs be large.  (In fact, we actually find that $R_{\rm TH}$ becomes slightly smaller than $R_{\rm Lag}$ at small masses.) Whereas the increase of $f$ at small masses is similar to that in Fig.~\ref{fig:LagrProfile_bestfit_WGWTH} of the main text, the decrease in $R$ here is larger than that in Fig.~\ref{fig:LagrProfile_bestfit_WGWTH}, perhaps because the parameter $A$ absorbs some of the effect there.

\section{ Dependence on halo finder}\label{sec:hf}
\subsection{ FoF linking length}\label{sec:ll_dependence}
The main text presented results using halos obtained with linking length $\ell=0.156$ times the interparticle separation.  Here we study how the Lagrangian profile $p_{\rm h}(r)$ depends on $\ell$, by comparing with results for $\ell=0.2$.  

\begin{figure}[!htb]
\centering
\includegraphics[width=\linewidth]{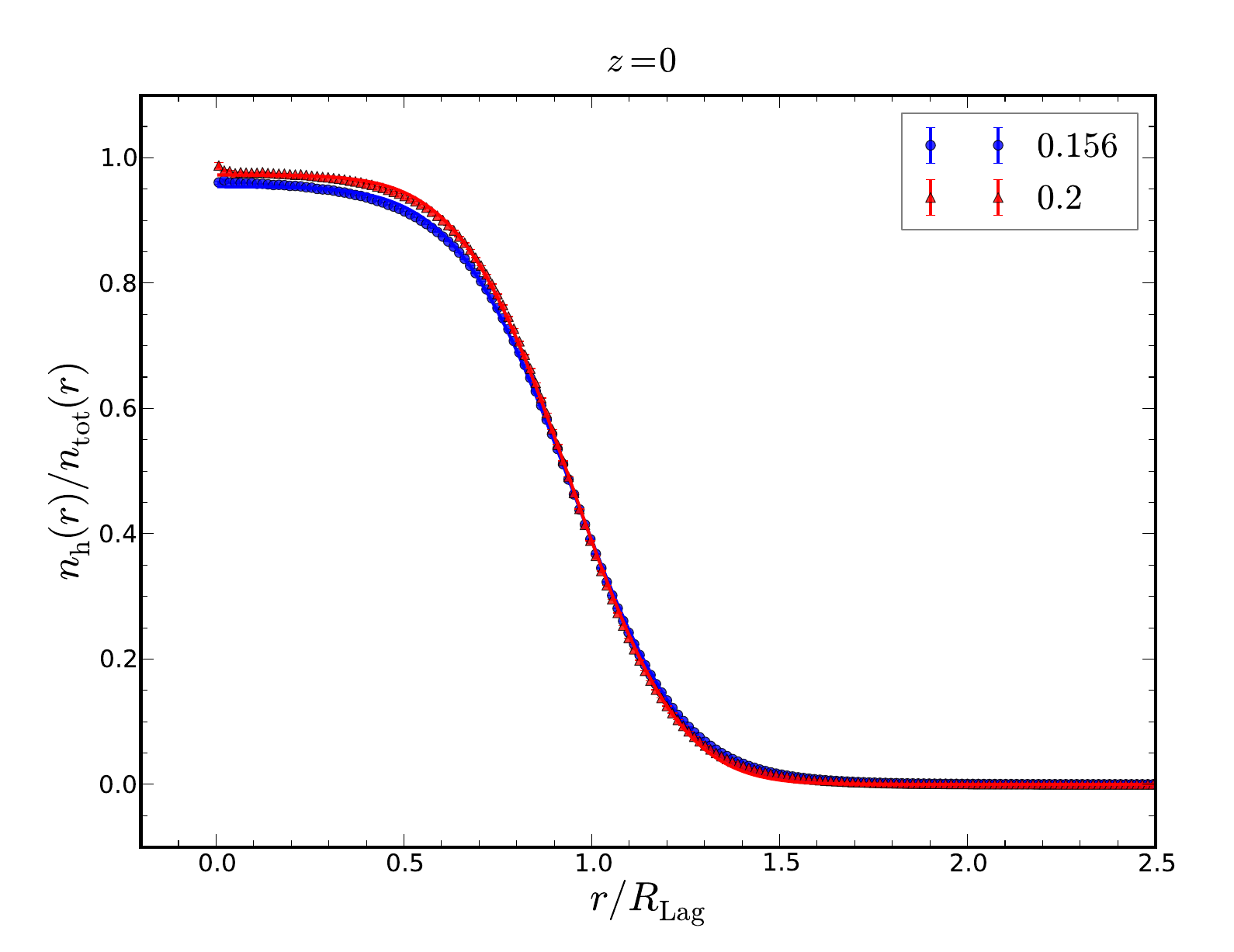}
\caption{ Lagrangian protohalo profiles of Eulerian halos identified with FoF link lengths $\ell=0.156$ (blue circles) and 0.2 (red triangles).  Both sets of profiles are for halos having mass $\sim 9.4 \times 10^{13} \Msun$, so the values of $R_{\rm Lag}$ are the same, but the objects themselves are not.  }
\label{fig:nh_ntot_ratio_ll_SameMass}
\end{figure}

Fig.~\ref{fig:nh_ntot_ratio_ll_SameMass} compares the protohalo profiles of objects identified in the $z=0$ outputs of the Carmen simulation using $\ell=0.15$ and $\ell=0.2$.  In both cases, we have chosen objects having mass $\sim 9.4 \times 10^{13} \Msun$ for which discreteness effects should not be an issue.  The objects identified using larger $\ell$ have profiles which are slightly more tophat-like, and which are slightly closer to unity at small $r$.  While this trend mimics the dependence on halo mass seen at fixed $\ell$, one should bear in mind that the $\ell=0.2$ objects would have had smaller masses in the $\ell=0.156$ catalog.  

%However, one should bear in mind that suppose a halo is identified with $b=0.156 $ to be of certain mass, when $b=0.2$ is applied, this halo will be classified to be a halo of larger mass. Thus if we compare the halos obtained with $b=0.156$ and $b=0.2 $ in a narrow mass range, we are in fact comparing different objects.

\begin{figure}
\centering
\includegraphics[width=\linewidth]{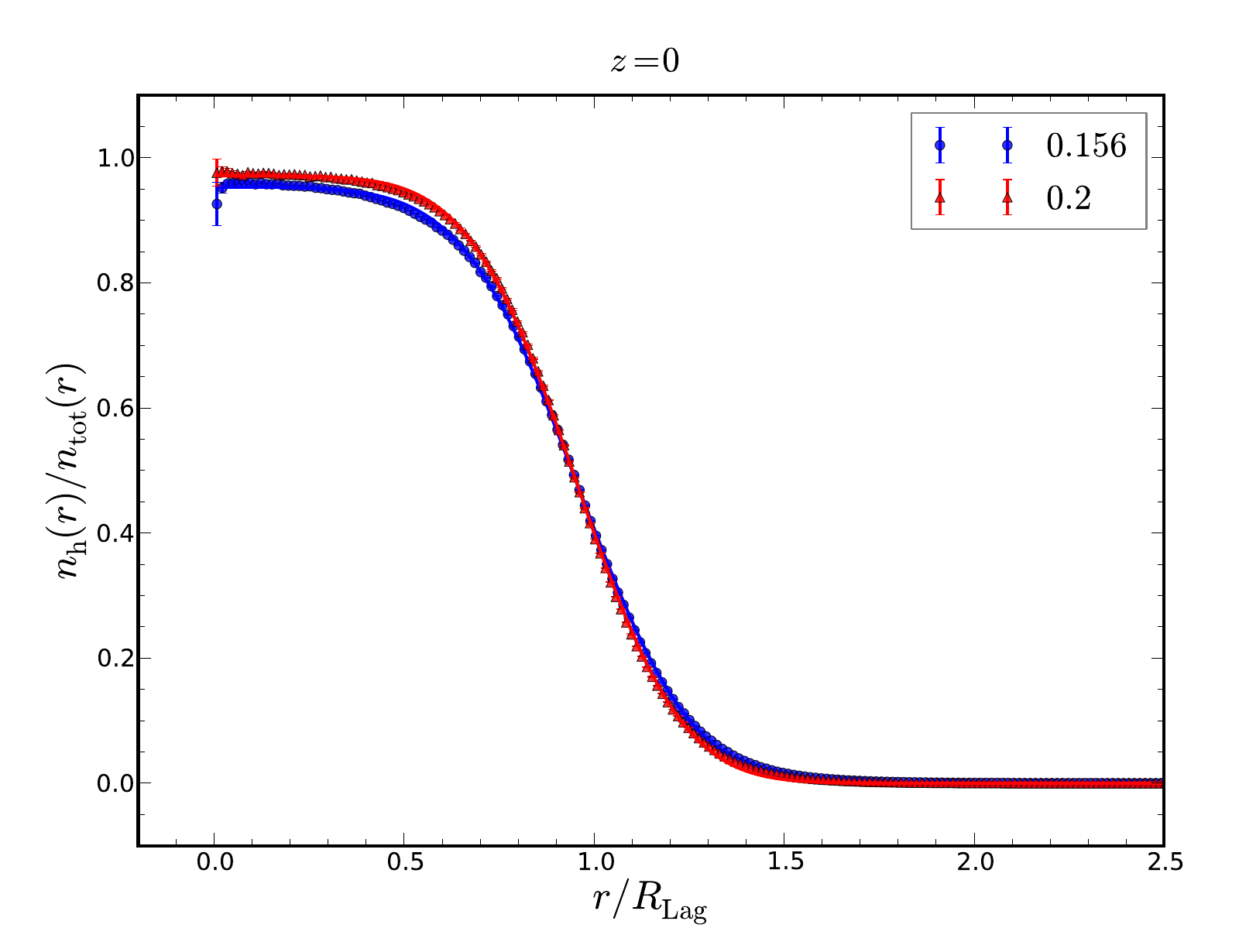} 
\caption{ Similar to Fig.~\ref{fig:nh_ntot_ratio_ll_SameMass} except in this plot, the samples are chosen such that the halos are the same in both cases.  However, the $\ell=0.156$ halo masses are $\sim 9.6 \times 10^{13} \Msun $, while the $\ell=0.2$ masses are $1.1 \times 10^{14} \Msun$, so the values of $R_{\rm Lag}$ differ by factor of about 1.06. }
\label{fig:nh_ntot_ratio_ll_SameHalo}
\end{figure}

We now select the samples such that we are comparing the same halos. We search the $\ell=0.2$ halos and select those that match the center of mass positions of $\ell=0.156$ halos. For the $\ell=0.156$ halos with masses $\sim 9.6 \times 10^{13} \Msun$, the corresponding $\ell=0.2$ halos have masses that are about 20\% larger:  $\sim 1.1 \times 10^{14} \Msun $.  Fig.~\ref{fig:nh_ntot_ratio_ll_SameHalo} compares their profiles, scaled by their respective $R_{\rm Lag}$ values (which differ by factor of about 1.06).  The results are  very similar to those in Fig.~\ref{fig:nh_ntot_ratio_ll_SameMass}:  larger $\ell$ results in a more compact protohalo profile.  Note that, had we rescaled both profiles by the same factor, say $R_{\rm Lag}$ of the $\ell=0.156$ masses, then the profiles would mainly have differed at small $r$, with the larger $\ell$ having larger $p_{\rm h}$.  This suggests that the difference from unity at small $r$ is real:  these are particles which were initially close to the protohalo center, but received large virial kicks, so in the $z=0$ snapshot they happen to lie beyond the boundary identified by the smaller linking length.
%Unfortunately, this does not imply that particles which are bound to the halo are better identified using a larger $\ell$.  E.g., the `halos' identified with $\ell=0.3$ would be substantially more massive; the associated protohalos would have , larger 

\subsection{ FoF vs SO halo finder  }\label{sec:FOFvsSO}
The main text used halos identified using an FoF halo finder with $\ell=0.156$.  Here we compare with results based on spherical overdensity (SO) halos that are $200\times$ the critical density and were identified using the public halo finder AHF \cite{Gill:2004km, Knollmann:2009pb}. 

\begin{figure}
\centering
\includegraphics[width=\linewidth]{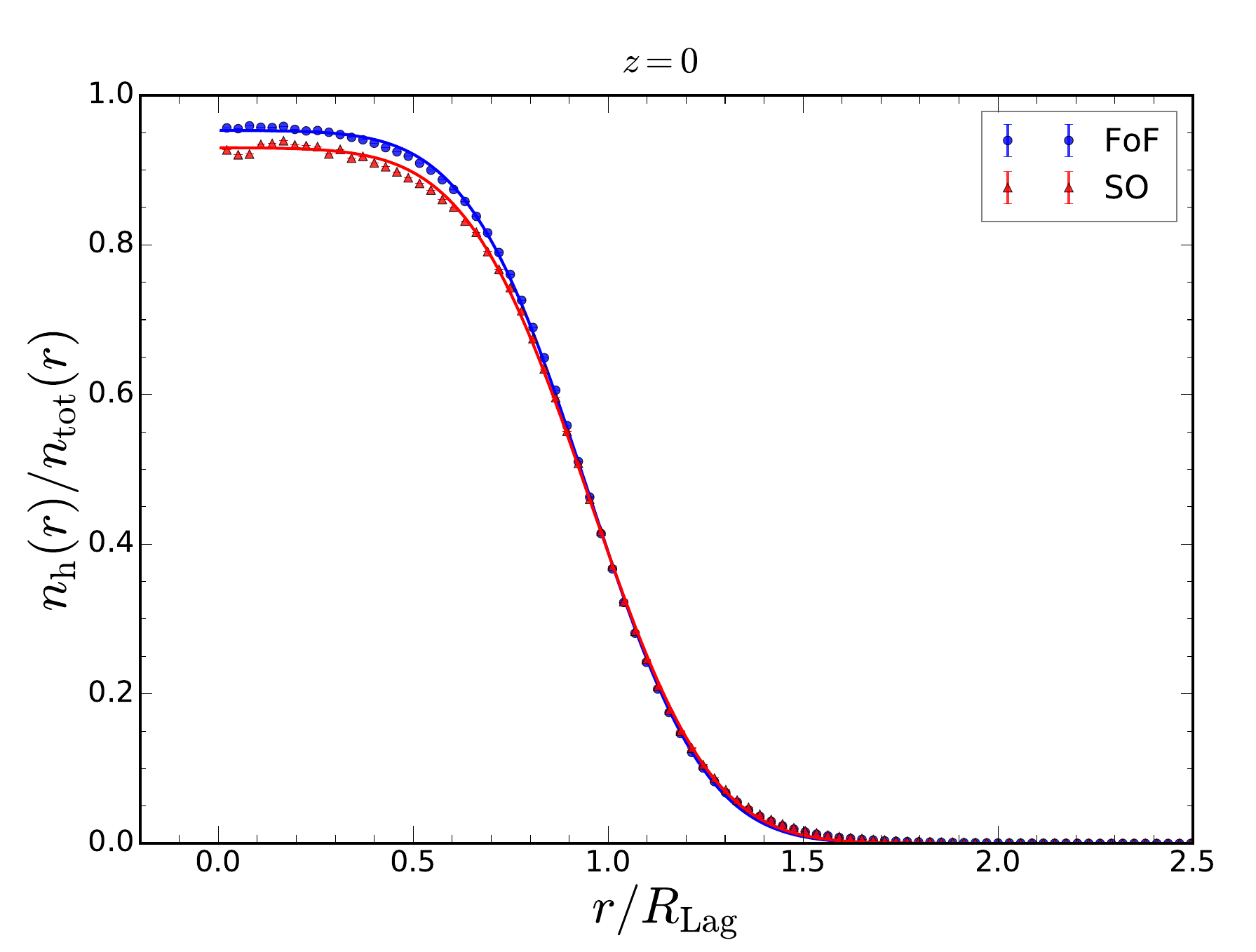}
\caption{ Lagrangian halo profile obtained from $\ell=0.156$ FoF (blue circles) and $200\rho_c$ SO (red triangles) halos identified at $z=0$ in the Carmen simulations.  Both FoF and SO masses are $\sim 9.4 \times 10^{13} \Msun$. Smooth curves show the result of fitting Eq.~\eqref{eq:WG*THrealspace} to these profiles. }
\label{fig:nh_nbar_r_halofinder_compare}
\end{figure}

Fig.~\ref{fig:nh_nbar_r_halofinder_compare} shows $p_{\rm h}$ obtained from the $\ell=0.156$ FoF (circles) and SO (triangles) halos identified at $z=0$ in the Carmen simulations having masses of $9.4 \times 10^{13} \Msun$.  At small $r$, the SO profile has noticeably smaller $p_{\rm h}$ than the FoF one, suggesting that the nonspherical shapes which FoF allows permit one to identify a greater fraction of the particles that are bound to the halo.  However, this is redshift-dependent. At $z=0.97$, the average SO protohalo profile is slightly more compact than the FoF one.  The smooth curves show that, like FoF protohalos, the SO protohalo profiles are also well fitted by our Eq.~\eqref{eq:WG*THrealspace}.

\section{Relation to spherical collapse and the collapse threshold measurements}
\label{sec:sc}
Conceptually, going beyond the tophat shape is subtle.  This is because the window function plays two roles.  In theory calculations, it smooths the initial overdensity field.  In the other, which is close to the way measurements are made, it is a weighting function which selects a fraction of the local dark matter particles that go on to become members of the Eulerian halo later on.  For a tophat, with a determinisitic relation between particle motions and distance from the center, these two roles appear to be the same, but for other filters, or when there is some stochasticity, they need not be.  This has two implications, which we use the excursion set approach to illustrate.

\subsection{Smoothing window versus density profile}\label{sec:pr}
In the main text, $p(r)$ was obtained using the particle distribution.  Therefore, it is tempting to associate it with a density profile, and hence to ask how well the spherical evolution model describes its evolution.  We now argue that, in this respect, it really is better to think of this distribution as a smoothing window, $W_{\rm eff}$, and not as a profile.

Specifically, it is tempting to define the mean overdensity inside $r$ as
\beq
\bar{ \eta }(\le r) = \frac{ 3 \int_{0}^r  dr'\, r'^2\, n_{\rm h}(r')   }{  r^3 \bar{n}_{\rm m}  } - 1 .
\eeq
However, since $n_{\rm h}=p_{\rm h}\,\bar{n}_{\rm m}$ and $p_{\rm h}\le 1$, the quantity above is guaranteed to be negative.  Therefore, if we were to use this $\bar{ \eta }$ as a proxy for overdensity in the spherical evolution model, we would conclude that the protohalo patch would not shrink and collapse (in comoving coordinates).  This is a wrong answer, of course, since the particles were identified precisely because they did form a collapsed object.  

The error arises because the argument above does not account for the fact that the particles coming from further away must have had higher infall velocities, and the naive application of the spherical model argument has ignored this.  It is straightforward to include the fact that the initial velocities are perturbed from the initial Hubble flow in the spherical model analysis.  But instead of doing so directly, we think the discussion which follows makes a similar point.  

First, define 
\beq
 \bar{ \delta }_{\rm Lag}
%  = \int_{0}^\infty  dr'\, 4 \pi  r'^2\, \delta_{\rm m}(r')\, W_{\rm eff}(r').
  = \int d \mb{x}'  \delta_{\rm m}( \mb{x}' )\, W_{\rm eff}(\mb{x}').
\eeq
This expression represents a smoothing of the initial overdensity fluctuation field centered on the origin (in this case, the center of the protohalo patch).  However, the continuity equation relates the overdensity to the divergence of the velocities:  so the average over $\delta_{\rm m}$ above is like averaging over the velocity divergence.  It is this smoothed overdensity which should be inserted into the spherical evolution model.  In this respect, $W_{\rm eff}\le 1$ in the expression above is a simple way to account for the fact that some particles fall onto the protohalo center and others escape; i.e., it is the motions, the displacements from the initial positions, which matter.  The role of $W_{\rm eff}$ is to only count those particles in the initial conditions whose displacements will bring them into the final collapsed object. It does not do this exactly, because the particles which do end up in the final object did not have the average infall speeds, whereas the expression above is assuming they did.  If the distribution around the mean is narrow, this will be a good approximation; otherwise, we expect $\bar{ \delta }_{\rm Lag}$ to underestimate the true effective overdensity.  This is what motivates the last of our three choices in Fig.~\ref{fig:mfn_esp}:  whereas using  $\delta_{\rm TH}$ with $W_{\rm TH}$ and $\delta_{\rm eff}$ with $W_{\rm eff}$ are natural choices, using $\delta_{\rm TH}$ with $W_{\rm eff}$ is a crude way of accounting for the fact that if $W_{\rm eff}$ is picking out the particles with the largest infall speeds, then their associated effective overdensity will be larger than $\delta_{\rm eff}$. 

Previous work has shown that not only is $\bar{\delta}_{\rm Lag}$ greater than zero, but when a tophat smoothing window is used, then $\bar{\delta}_{\rm Lag}\ge\delta_{\rm sc}$ \cite{ShethMoTormen2001,DespaliTormenSheth2013}.  Therefore, we expect it will also be of order $\delta_{\rm sc}$ when $W_{\rm eff}$ is used, and indeed it is (see \cite{LagBiasPaper} and Fig.~\ref{fig:scatter_dist_lognormal_fit}).  This is the sense in which $W_{\rm eff}$ can be thought of as playing two distinct roles:  it is both a fraction of particles, and a smoothing filter which should be applied to the Lagrangian overdensity fluctuation field.  

The analysis above suggests that protohalo particles are those for which $v/fHr \sim \delta_{\rm sc}/3$, a point we flesh out in slightly more detail in the next subsection.

\subsection{Infall speeds and protohalo patches}\label{sec:fuzzy}
In the idealized spherical collapse calculation, the initial protohalo is a tophat, with a determinisitic relation between particle motions and distance from the center.  As a result, the requirement that the particles fall towards to the protohalo center of mass and arrive there by the present time, and the requirement that the enclosed density within the patch have a certain critical value, are the same.  In reality, things are more complicated.  

To see why, suppose that we are centered on a patch in Lagrangian space around which the overdensity, smoothed with a tophat of radius $R$, equals $\delta_{\rm c}$, and that the overdensity on the next larger smoothing scale is less than this.  If we average over many such patches, we can define the mean overdensity as a function of distance $r$ from the center.  The Fourier transform of $\langle \delta(\le r)|\delta_0,\delta_0'\rangle$, the average overdensity within $r$, is the quantity which played a major role in the main text (although here the smoothing filter is obviously $W_{\rm TH}$).  Our notation reflects the fact that the value of $\delta$ and its derivative at the origin are constrained.

Similarly, at each $r$, there will be a distribution of radial infall speeds.  Let $p(v_r|\delta_0,\delta_0')$ denote this distribution.  If the initial fluctuation field was Gaussian, then this distribution is Gaussian with a mean $\langle v_r|\delta_0,\delta_0'\rangle = fHr\,\langle \delta(\le r)|\delta_0,\delta_0'\rangle/3$ which depends on the value of the constraints, and a variance $\sigma^2(v_r|\delta_0,\delta_0')$ whose value depends on the fact that the enclosed overdensity and its slope were constraints, but does not depend on $\delta_0$ and $\delta_0'$ themselves.

The particles which will end up in the final halo are those with sufficiently large radial infall speeds.  At each $r$, this is the fraction 
\beq
 p(r) = \int_{v(r)}^\infty dv_r \, p(v_r|\delta_0,\delta_0')
\eeq
where $v(r)\sim fH(r-r_{\rm vir})\sim fH(r - R_{\rm Lag}/5)$.  Clearly, the more distant particles must have larger infall speeds so as to arrive within $r_{\rm vir}$ by the present time.  E.g., in the Zeldovich approximation,
\beq
 p(r) = \frac{1}{2} {\rm erfc}\left(\frac{r - r_{\rm vir}-r\,\langle \delta(\le r)|\delta_0,\delta_0'\rangle/3}{\sqrt{2}\sigma_{v|\delta,\delta'}/H}\right),
 \eeq
so that
\beq
 p(R_{\rm Lag}) =
 \frac{1}{2} {\rm erfc} \left(\frac{4/5 - \delta_c/3}{\sqrt{2}\sigma_{v|\delta,\delta'}/fHR_{\rm Lag}}\right) \sim 1/2.
\eeq
But at $r\gg R_{\rm Lag}$, the term in brackets tends to $fHr/\sigma_{v}$ so $p(r)\to 0$.  This shows that:
even if $\delta_0$ was identified using a real space tophat of scale $R_{\rm Lag}$, not all the particles within $R_{\rm Lag}$ will make it into the final halo;
 and some of the particles which are initially beyond $R_{\rm Lag}$ will end up in the final halo.
As a result, $p(r)$ will be more extended than a tophat, in qualitative agreement with what we found in the main text.  I.e., in this picture, one thinks of protohalo particles as those which satisfy a tophat in `infall time' towards the protohalo center, and this need  {\em not} correspond to a tophat in the initial spatial distribution.

\bibliography{Lagr_Weff}

\end{document}